\documentclass[aps,prx,twocolumn,longbibliography, footinbib]{revtex4-2}

\usepackage{soul}
\usepackage[dvipsnames]{xcolor}
\usepackage{amsmath}
\usepackage{amsfonts}
\usepackage{graphicx}

\usepackage{amssymb}
\usepackage{amsbsy}
\usepackage{color}

\newcommand{\bra}[1]{\langle{#1}|}
\newcommand{\ket}[1]{|{#1}\rangle}

\newcommand{\beq}{\begin{equation}}
\newcommand{\eeq}{\end{equation}}
\newcommand{\bea}{\begin{eqnarray}}
\newcommand{\eea}{\end{eqnarray}}

\usepackage{hyperref}
\hypersetup{
	colorlinks=true,
	citecolor=blue,
	linkcolor=blue,
	urlcolor=blue}

\begin{document}

\title{Dissipation as a Resource: Synchronization, Coherence Recovery, and Chaos Control}

\author{Debabrata Mondal$^1$, Lea F. Santos$^{2}$, S. Sinha$^1$}
\affiliation{$^1$Indian Institute of Science Education and Research-Kolkata, Mohanpur, Nadia-741246, India\\
$^2$Department of Physics, University of Connecticut, Storrs, Connecticut 06269, USA \\	
}

\begin{abstract}
Dissipation is commonly regarded as an obstacle to quantum control, as it induces decoherence and irreversibility. Here we demonstrate that dissipation can instead be exploited as a resource to reshape the dynamics of interacting quantum systems. Using an experimentally realizable Bose-Josephson junction containing two bosonic species, we demonstrate that dissipation enables distinct dynamical behaviors: synchronized phase-locked oscillations, transient chaos with long-time coherence recovery, and steady-state chaos. The emergence of each behavior is determined by experimentally tunable parameters. At weak interactions, the two components synchronize despite dissipation, exhibiting long-lived coherent oscillations reminiscent of a boundary time crystal. Stronger interactions induce a dissipative phase transition into a self-trapped regime accompanied by chaotic dynamics. Remarkably, dissipation regulates the lifetime of chaos and enables the recovery of coherence at long times. By introducing a controlled tilt between the wells, transient chaos can be converted into persistent steady-state chaos. We further show that standard spectral diagnostics fail to distinguish between the two chaotic regimes, revealing that spectral statistics primarily reflect short-time instability. These results establish dissipation as a powerful tool for engineering dynamical phases, restoring quantum coherence, and controlling the duration of chaotic behavior and information scrambling.
\end{abstract}
\maketitle

\section{INTRODUCTION}

Open interacting quantum systems represent a vibrant frontier in contemporary physics, where the competition between coherent dynamics, interactions, and dissipation gives rise to a wealth of non-equilibrium phenomena~\cite{Mivehvar2021, Carusotto2013, Diehl2010, Ritsch2013}. 
Traditionally regarded as a detrimental source of decoherence and noise, dissipation has recently emerged as a powerful control mechanism, capable of stabilizing nontrivial dynamical states and enabling phases of matter that have no counterpart in closed systems.  Examples include the dissipative stabilization of strongly correlated steady states~\cite{Poyatos1996, Verstraete2009, Diehl2008, Tomita2017, Harrington2022}, the emergence of non-Hermitian topological phases \cite{Ashida2020,Bergholtz2021}, and the formation of time-crystals~\cite{Kessler2021, Kongkhambut2022, Wu2024, Souza2023, Russo2025, Dutta2025},  
where time-translation symmetry is broken spontaneously and oscillations persist for long times. In driven-dissipative quantum systems, synchronization can emerge as phase or frequency locking, either among interacting subsystems or with an external drive~\cite{Laskar2020, Li2025SciAdv, Walter2014, Mari_PRL_2013, Roulet_PRL_2018}. These discoveries not only deepen our understanding of non-equilibrium quantum dynamics but also open new avenues for precision metrology and quantum control~\cite{Verstraete2009, Pezze2018, Lesanovsky_Time_crystal_2024, WisemanMilburn2010, Sekatski2017}.  

Dissipation also plays a central role in critical phenomena and chaotic dynamics. In dissipative phase transitions, the steady state of an open system changes non-analytically as system parameters are varied, extending the notion of equilibrium criticality to Liouvillian dynamics~\cite{Minganti_2018, Fitzpatrick_2017, Hemmerich_PNAS_2015}. In parallel, studies of dissipative quantum chaos~\cite{Haake_GHS, Beneti_2005, Vivek_ADD, Minganti, Deb_TCD, Prosen_PRX, Prosen_PRL, Kulkarni_Dicke, Lea_Santos_GHS, Robnik_225} raise fundamental questions about the effects of loss on chaotic behavior and about the extent to which spectral correlations can serve as signatures of chaos~\cite{Lea_Santos_GHS, Robnik_225, Deb_OADM_2026}.

These developments indicate that dissipation alters not only steady states but also the structure of quantum dynamics, influencing critical behavior, chaos, and information scrambling in ways that remain actively under investigation. Understanding how dissipation controls the emergence, stability, and temporal structure of complex quantum dynamics has therefore become essential, both from a fundamental perspective and for the development of robust quantum technologies~\cite{Verstraete2009, Diehl2008, Tomita2017, Harrington2022, Lesanovsky_Time_crystal_2024}.
At the same time, state-of-the-art experimental platforms -- ranging from ultracold atoms in optical lattices~\cite{rev0, rev1} and superconducting circuits~\cite{ Steven_Girvin_1, Houck} to trapped ions~\cite{ Bollinger_2012, A_M_Rey_2016} -- now offer unprecedented control over both Hamiltonian parameters and dissipation mechanisms, creating an ideal setting to explore dissipation as an active element of quantum dynamics.

Here, we show that dissipation can be harnessed as a control knob to reshape the dynamics of interacting quantum systems. We focus on a two-component  Bose-Josephson junction (BJJ) \cite{Mondal_TCBJJ_2022}, consisting of a double well potential populated by two distinguishable bosonic components, where dissipation arises from environment-induced incoherent hopping between the wells. This system is experimentally realizable and admits a transparent classical-quantum correspondence. It can also be equivalently formulated as a model of two large collective spins (two coupled tops), a setting relevant across cold-atom, solid-state, and quantum-optical platforms.

By combining classical phase-space analysis with quantum trajectory simulations, we identify qualitatively distinct dynamical regimes stabilized by dissipation: synchronized phase-locked oscillations, dissipative phase transition to a self-trapped regime, transient chaos accompanied by long-time coherence recovery, and steady-state chaos. Figure~\ref{Fig1} provides a schematic representation of these main findings. The transitions between these regimes are governed by simple and experimentally tunable parameters, including the interaction strength and a controllable tilt between the wells. 

For weak interactions, the population imbalances and relative phases of the two bosonic components lock, resulting in long-lived synchronized oscillations even in the presence of dissipation and the absence of external driving.
This phase locking dynamically reduces the number of effective degrees of freedom and gives rise to an emergent constant of motion. The resulting persistent oscillatory state is closely analogous to a boundary time crystal~\cite{Boundary_time_crystal}.
 
Increasing the interaction strength beyond a critical value destabilizes the synchronized oscillatory phase and drives the system into a self-trapped regime, in which atoms preferentially occupy one of the wells. Crossing this critical point marks a 
dissipative phase transition, beyond which dissipation gives rise to a stable attractor that governs the long-time dynamics and renders the steady state insensitive to initial conditions.

Beyond the critical interaction strength, the system also enters a regime of chaotic dynamics. Nevertheless, we can employ  dissipation to control the duration of this behavior, so that information scrambling and dephasing are only temporary. At long times, dissipation drives the system toward the stable attractor, leading to the recovery of coherence. These results establish dissipation as a powerful mechanism for regulating chaos and preserving long-time quantum coherence.

Remarkably, introducing a tilt between the two wells destabilizes the attractor, converting transient chaos into steady-state chaos. In this regime, coherence is not recovered and information scrambling persists at long times. The transition from transient to steady-state
chaos reveals how dissipation can control not only the onset but also the persistence of chaotic dynamics in open
quantum systems.  Interestingly, the Liouvillian spectral statistics do not distinguish between these regimes: both exhibit Ginibre correlations, indicating that non-Hermitian universality alone is insufficient to differentiate transient from steady-state chaos. 

Taken together, our results establish a unified dynamical framework in which dissipation acts as a  versatile control resource, capable of stabilizing coherent motion, inducing non-equilibrium phase transitions, and regulating chaotic dynamics. Dissipation can be engineered to control not only whether chaos emerges, but how long it persists, opening new avenues for dissipation-based control of complex quantum dynamics in experimentally accessible platforms.

\section{RESULTS}
After introducing the open two-species BJJ, we present our results on its dynamical phases, highlighting how interactions and dissipation jointly shape synchronization, self-trapping, and chaos. We further demonstrate how a tilted potential provides a control knob to tune chaotic dynamics from transient to steady-state behavior.

\begin{figure}[t]
\centering    \includegraphics[width=\linewidth]{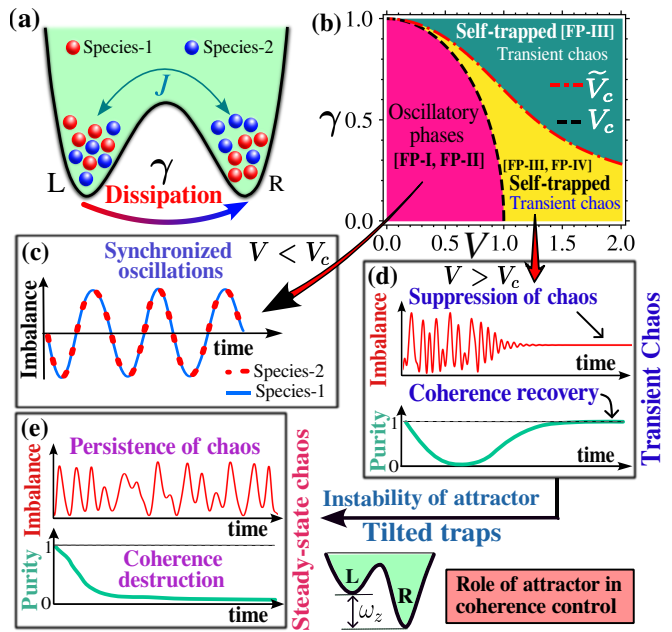}
\caption{{\bf Dissipation-controlled dynamical regimes of an open two-component Bose-Josephson junction}.
(a) Schematic of the system with coherent tunneling, interspecies interactions, and incoherent hopping.
(b) Phase diagram in the interaction ($V$)–dissipation ($\gamma$) plane.
(c) Weak interactions yield synchronized oscillations even in the presence of dissipation.
(d) Stronger interactions lead to transient chaos, where dissipation suppresses long-time scrambling and restores coherence due to the presence of a stable attractor.
(e) A tilt destabilizes the attractor, producing steady-state chaos with persistent decoherence.
}
\label{Fig1}
\end{figure}

\textbf{Quantum and classical model}: 
We consider a two-component BJJ with equal boson populations $N$ in a double-well potential, described by the Hamiltonian
\begin{equation}
\hat{\mathcal{H}}=-\frac{J}{2} \sum_{i=1}^2\left( \hat a^\dagger_{iL}\hat a_{iR} + \hat a^\dagger_{iR}\hat a_{iL} \right)+
\frac{V}{N} \left( \hat n_{1L}\hat n_{2L} + \hat n_{1R}\hat n_{2R} \right)
\label{BJJ}
\end{equation}
where $L, R$ label the left and right wells, $i $ denotes the two bosonic species, operator $\hat{a}_{iL(R)}$  annihilate a boson of species $i$ in well $L$ $(R)$, and $\hat{n}_{iL(R)}= \hat{a}^\dagger_{iL(R)} \hat{a}_{iL(R)}$ is the number operator. 
The first term describes the hopping between the two wells with amplitude $J$, while the second term accounts for the repulsive interspecies interaction of strength $V$. We set $\hbar=1$ and measure energy (time) in units of $J\ (1/J)$.

Motivated by the experimental realization of environment-mediated incoherent atomic hopping in optical lattices ~\cite{Labouvie_2016, Lohse2016}, we include an incoherent hopping process from the left to the right well with rate $\gamma$, which conserves the total number of atoms of each species. This process can be activated by
noise or thermal fluctuations when the left cavity is coupled to a bath~\cite{Garbe2024}. This dissipative process is described by the jump operators $\hat{\mathcal{O}}_i = \sqrt{2\gamma/N}\hat{a}_{iR}^{\dagger}\hat{a}_{iL}$ leading to the Lindblad master equation for the total density matrix
\begin{eqnarray}
	&\dot{\hat{\rho}}&\,\, =\!-i[\mathcal{\hat{H}},\hat{\rho}]\!+\!\sum_{i=1,2}  \left(\hat{\mathcal{O}}_i\rho \hat{\mathcal{O}}_i^{\dagger}-\frac{1}{2} \hat{\mathcal{O}}_i^{\dagger}\hat{\mathcal{O}}_i\rho-\frac{1}{2}\rho \hat{\mathcal{O}}_i^{\dagger}\hat{\mathcal{O}}_i\right)\!\!. \qquad
	\label{Master_equation}
\end{eqnarray}
A schematic representation of the open BJJ is given in  Fig.~\ref{Fig1}(a).

We can recast the model in terms of collective spin operators using the
Schwinger boson representation. Defining $\hat S_{ix}=\frac{1}{2}\!\left(\hat a^\dagger_{iL}\hat a_{iR}+\hat a^\dagger_{iR}\hat a_{iL}\right)$ and $\hat S_{iz}=\frac{1}{2}\!\left(\hat n_{iL}-\hat n_{iR}\right)$, each bosonic species maps onto a collective spin of magnitude $S=N/2$. Under this mapping, the BJJ Hamiltonian in Eq.~(\ref{BJJ}) reduces to the coupled-top model~\cite{D_mondal_CT_2022},
\begin{eqnarray}
\hat{\mathcal{H}}=-J\hat{S}_{1x}-J\hat{S}_{2x}+\frac{V}{S}\hat{S}_{1z}\hat{S}_{2z},
\label{Hamiltonian_CT}
\end{eqnarray}
while the dissipative dynamics is described by the jump operators $\hat{\mathcal{O}}_i = \sqrt{\frac{\gamma}{S}} \hat{S}_{i-}$, corresponding to collective spin decay~\cite{Exactness_2}.  This spin formulation provides a compact description of the system  and establishes a direct connection with a broad class of collective models studied in cold-atom, trapped-ions, and molecular-magnet platforms~\cite{Deutsch_2007, Sohini_Ghose_2009, Oberthaler_1, Oberthalar_2, KCT_Magnetic_cluster_1, KCT_Magnetic_cluster_2}.

The model admits a well-defined classical limit for a large atom number $N\gg1$ (equivalently, large spin magnitude $S\gg1$). In this limit, the dynamics of each bosonic species can be described in terms of macroscopic mode amplitudes
$\alpha_{iL(R)}=\langle \hat a_{iL(R)}\rangle$,
which can be parametrized as
$\alpha_{iL(R)}=\sqrt{n_{iL(R)}}\,e^{i\theta_{iL(R)}}$, where $n_{iL(R)}=\langle \hat n_{iL(R)}\rangle$ denotes the average number
of atoms in the left (right) well and $\theta_{iL(R)}$ is the corresponding condensate phase. We then introduce the population imbalance between the two wells and the
relative phase of the collective wave function
\begin{equation}
z_i=\frac{n_{iL}-n_{iR}}{N}, \qquad
\phi_i=\theta_{iR}-\theta_{iL},
\label{Eq:CanonicalPair}
\end{equation}
which form a canonically conjugate pair. In the spin representation, the normalized spin expectation values
$s_{i\mu}=\langle \hat S_{i\mu}\rangle/S$, with $\mu=x,y,z$, behave as classical variables. Each species is thus represented by a classical spin vector
$
\vec{s}_i=\{\sin \vartheta_i \cos \phi_i,\sin\vartheta_i\sin\phi_i,\cos \vartheta_i \}
$
on the Bloch sphere, where the polar angle $\vartheta_i$ is related to the population imbalance by
$
z_i=\cos\vartheta_i\equiv s_{iz}
$.
The resulting equations of motion are 
\begin{subequations}
\begin{eqnarray}		
\dot{z}_i &=&-J\sqrt{1-z_i^2}\sin{\phi_i}-\gamma(1-z_i^2) ,\\
\dot{\phi}_i&=&\frac{Jz_i\cos{\phi_i}}{\sqrt{1-z_i^2}}+ V z_{\bar{i}},
\end{eqnarray}
\label{dissipative_EOM}	
\end{subequations}
where $\bar{i} \neq i$. 

In the quantum description, for each species, the population imbalance is represented by the operator $\hat{z}_i=(\hat{n}_{iL}-\hat{n}_{iR})/N$ and the relative phase between the wells can be obtained by appropriately constructing phase operators. The expectation values of observables and the density matrix are computed using an ensemble of quantum trajectories within the stochastic wave-function formalism (see Methods).

Using complementary quantum and classical descriptions, we now analyze the distinct dynamical phases that emerge as the interaction strength $V$ is increased from weak values. These regimes are summarized in the classical phase diagram shown in  Fig.~\ref{Fig1}(b).

\textbf{Synchronized oscillatory dynamics:}
We first focus on the weak-coupling regime, $V<V_c=\sqrt{J^2-\gamma^2}$, where the open two-component BJJ exhibits a coherent oscillatory phase whose frequency depends on the dissipation strength. In this regime,  coherent oscillations of the population imbalance $z_i$ and relative phase $\phi_i$ persist for both condensate species for long times despite the absence of external driving. This behavior is closely related to boundary time-crystal
phenomenology~\cite{Boundary_time_crystal}, where persistent temporal oscillations arise in open quantum systems governed by time-independent Liouvillians, reflecting the spontaneous breaking of continuous time-translation symmetry.

\begin{figure}[t]
\centering
\includegraphics[width=\linewidth]{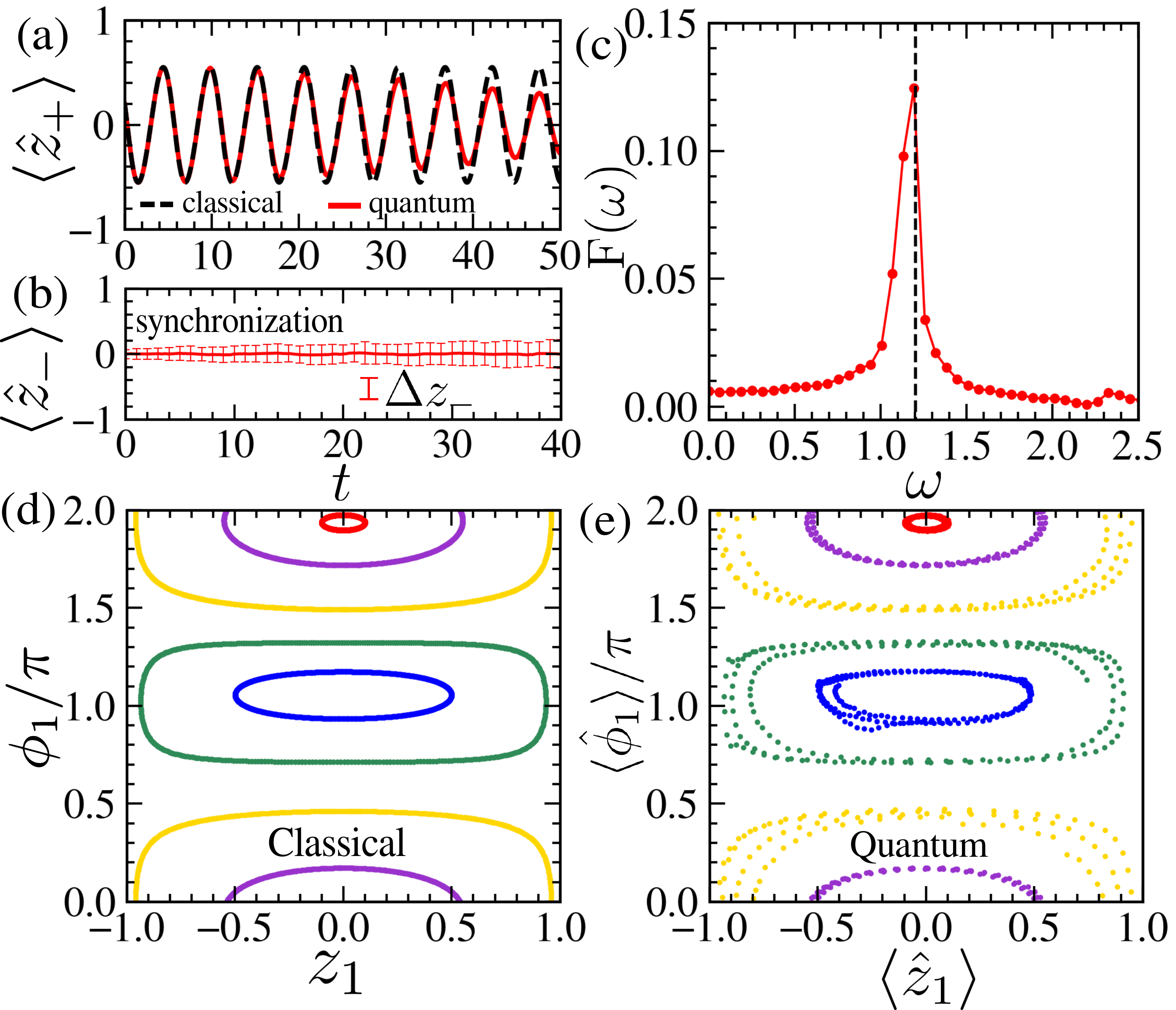}
\caption{{\bf Coherent synchronized oscillations.} (a,b) Dynamics of the (a) sum $\langle \hat{z}_+\rangle$ and (b) difference $\langle \hat{z}_-\rangle$ of population imbalances of the two species. The standard deviation $\Delta z_{-}$ is plotted in (b) as error bars. (c) The Fourier spectrum $F(\omega)$ of $\langle \hat{z}_{+}(t)\rangle$ reveals a single frequency of oscillation. The vertical dashed line represents the frequency of the corresponding classical orbit.
(d) [(e)]: Classical [quantum] oscillatory dynamics on the phase-space portrait of the conjugate plane $z_1-\phi_1$; in (e), the trajectories are plotted within $t\in [0,20]$. We choose $V=0.5$, $\gamma=0.2$, and $S=50$. In this and the rest of the figures, we choose $J=1$.
}
\label{Fig2}
\end{figure}

The emergence of the synchronized periodic oscillations is illustrated in Figs.~\ref{Fig2}(a)-(b) by comparing quantum and classical evolutions. The quantum dynamics is obtained from ensembles of quantum trajectories using the stochastic wave-function formalism~\cite{Daley_2014} for a relatively small atom number $N$ (see Methods). Both species are initiated in identical coherent states $\ket{z_i,\phi_i}$, corresponding to parallel spin configurations in the spin representation. Figure~\ref{Fig2}(a) shows  the time evolution of the symmetric population imbalance, 
\begin{equation}
 \langle \hat{z}_{+}(t) \rangle = (\langle \hat{z}_1 (t)\rangle + \langle \hat{z}_2 (t)\rangle)/2 ,  
\end{equation}
which captures the collective (synchronized) dynamics of the two species. The persistence of periodic oscillations in $\langle \hat{z}_{+}(t) \rangle$ demonstrates the emergence of a long-lived oscillatory state. Figure~\ref{Fig2}(b) displays the antisymmetric imbalance, 
\begin{equation}
 \langle \hat{z}_{-} (t)\rangle = (\langle \hat{z}_1 (t) \rangle - \langle \hat{z}_2 (t) \rangle)/2,   
\end{equation}
which quantifies deviations from synchronization. As seen in the figure, $\langle \hat{z}_{-} (t)\rangle$ remains vanishingly small throughout the evolution, indicating robust synchronization between the two species. Although quantum fluctuations are always present,  the error bars representing the standard deviation $\Delta z_{-}$ remain small when averaged over quantum trajectories.
The Fourier spectrum of the oscillations of $\langle \hat{z}_{+} (t)\rangle$ is shown in Fig.~\ref{Fig2}(c). It exhibits a single frequency, reflecting the coherent collective oscillation of the synchronized junction. 

The analysis of the equations of motion in Eq.\eqref{dissipative_EOM} reveals that in the weak-coupling regime, the classical dynamics is governed by two stable fixed points (FPs) in phase space, 
\begin{eqnarray}
\bullet &&\,\,\text{FP-I:}\quad \{z_i^*=0,\phi_i^*=\sin^{-1}(\gamma/J)+\pi\},\nonumber\\
\bullet &&\,\,\text{FP-II:}\quad \{z_i^*=0,\phi_i^*=-\sin^{-1}(\gamma/J)\}\nonumber.	
\end{eqnarray}
Remarkably, despite the presence of dissipation, both fixed points are centers rather than attractors, which gives rise to stable oscillatory motion around them. Small-amplitude oscillations about these fixed points generate coherent oscillations with characteristic frequencies determined by the interaction and dissipation rate as
\begin{eqnarray}
\omega_{\pm} = \frac{1}{J}\sqrt{(J^2-\gamma^2)^{\frac{1}{2}}[(J^2-\gamma^2)^{\frac{1}{2}}\pm V]}	,
\label{Oscillation_frequency}
\end{eqnarray}
where $\omega_{+}$ is the frequency of the symmetric (in-phase) mode  and $\omega_{-}$ is the frequency of the antisymmetric (out-of-phase) mode. 

Figure~\ref{Fig2}(d) presents classical phase-space trajectories of a single species in the synchronized oscillatory regime, plotted in the $(z_1,\phi_1)$ plane. They are closed orbits indicating periodic oscillations around one of the stable fixed points, FP-I or FP-II. Figure~\ref{Fig2}(e) displays the corresponding quantum phase-space dynamics, obtained from the expectation values 
$(\langle \hat{z}_1 \rangle, \langle \hat{\phi}_1 \rangle)$ averaged over quantum trajectories. Remarkably, the quantum evolution reproduces classical periodic-orbit–like structures up to a finite timescale. Such quantum-classical correspondence demonstrates the robustness of synchronized oscillatory dynamics despite  the presence of quantum fluctuations for finite spin magnitude $S$.

To assess the intra-species coherence of the oscillatory dynamics, we evaluate the fluctuation of the relative phase between the left and right wells of each condensate, defined as $(\Delta \phi)^2 = \langle \hat{\phi}^2 \rangle - \langle \hat{\phi} \rangle^2$, where $\hat{\phi}$ denotes the quantum mechanical phase operator (see Methods). 
In the oscillatory regime, the phase fluctuations grow very slowly, indicating that each condensate preserves a high degree of phase coherence during the oscillatory motion (see Supplementary Information (SI) \cite{SM}). This confirms that, despite the presence of incoherent hopping, the dynamics for each species remains effectively coherent.

The stable fixed points FP-I and FP-II are phase biased, in the sense that the relative
phase satisfies $\phi_i^{*}\neq 0,\pi$. As a result, each species sustains a finite
steady atomic current flowing from the left to the right well,
\[
\mathcal{J}_i^{L\rightarrow R} \equiv
-\,\frac{d}{dt}\,\langle \hat n_{iL} \rangle
=
\frac{J}{2i}
\left\langle
\hat{a}^{\dagger}_{iR}\hat{a}_{iL}
-
\hat{a}^{\dagger}_{iL}\hat{a}_{iR}
\right\rangle
=
- J \langle \hat{S}_{iy} \rangle .
\]
This steady current 
is analogous to the DC Josephson effect in superconducting junctions~\cite{Levy2007}, where a constant phase bias supports a persistent current even in the absence of an applied voltage. Here, however, the phase bias and the associated current are dynamically selected by dissipation, making the steady current an intrinsically non-equilibrium feature of the open BJJ with no counterpart in the isolated system.

The synchronized motion in the BJJ is a dynamical phenomenon that originates from the exchange symmetry of the equations of motion between the two species, $1 \leftrightarrow 2$. In this regime, the two bosonic species dynamically lock both their population imbalances, $z_1 \approx z_2$, and their relative phases, $\phi_1 \approx \phi_2$. As a consequence of this phase locking, the dynamics becomes effectively confined to the symmetric subspace, reducing the problem to that of a single collective condensate degree of freedom. This reduction gives rise to an emergent constant of motion, so that the synchronized dynamics can be effectively described by a single-species Lipkin-Meshkov-Glick model (see SI~\cite{SM}). 

The choice of initial conditions plays an important role in the emergence of synchronized oscillatory dynamics. Perfect synchronization is obtained for symmetric initial states, where the two species evolve within the symmetric sector of the dynamics. Nevertheless, even when the initial conditions deviate from exact symmetry, the system continues to display robust oscillatory behavior. In this case, the dynamics becomes quasiperiodic, characterized by a small number of dominant frequencies rather than a single fundamental one. This behavior is reminiscent of continuous time quasicrystals~\cite{He2025PRX}. 

We now contrast the synchronized behavior described above with the strong-interaction regime. As discussed below and detailed in the SI~\cite{SM}, increasing the interaction strength beyond the critical value, $V>V_c$, destroys synchronization.


\textbf{Dissipative phase transition:} At the interaction strength $V_c=\sqrt{J^2-\gamma^2}$, the Josephson frequency $\omega_{-}$ vanishes, and for $V>V_c$, it becomes imaginary,  signaling an instability of the fixed points FP-I and FP-II. The critical value $V_c$ thus marks a transition from the synchronized oscillatory phase to a self-trapped regime characterized by finite population imbalance and a stable attractor. The transition line is indicated with a black dashed curve in Fig.~\ref{Fig1}(b) and can be understood from the classical analysis below.

\begin{figure}[t]
\centering
\includegraphics[width=\linewidth]{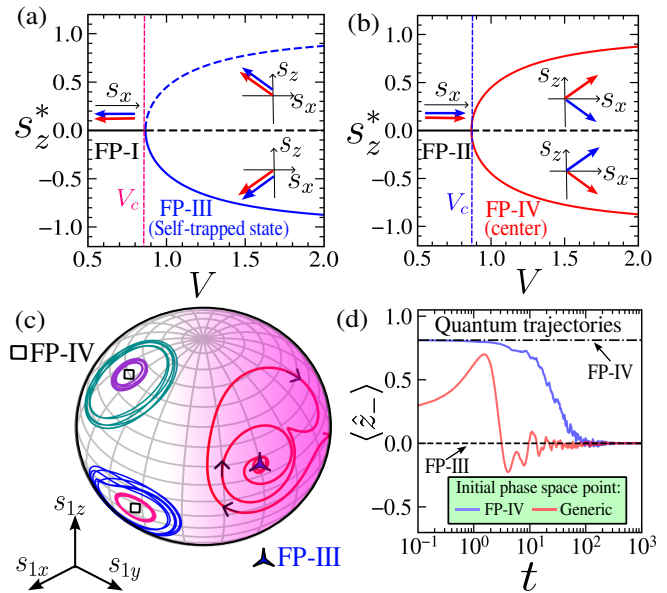}
\caption{{\bf Dissipative phase transition from synchronization to self-trapping.}
Bifurcation diagrams of $s_z^*$ corresponding to (a) transition from fixed point FP-I to FP-III and (b) from FP-II to FP-IV. Solid (dashed) lines indicate stable (unstable) fixed points.  (c) Trajectories on the Bloch sphere of one spin exhibiting oscillation around FP-IV and relaxation dynamics to the attractor FP-III.  (d) Quantum dynamics of $\langle \hat{z}_-\rangle$, averaged over trajectories, starting close to FP-IV (blue line) and at a generic state (red line). Both trajectories ultimately converge to FP-III.
We use $\gamma=0.2$ in all panels and $V=1.7$ in (c,d). For the quantum result in (d), $S=30$.
}
\label{Fig3}
\end{figure}

In the isolated case ($\gamma=0$), the fixed point FP-I undergoes a pitchfork bifurcation at $V=V_c$, leading to a symmetric self-trapped configuration in which both species accumulate in the same well, $z_1^*=z_2^*=\pm\sqrt{1-J^2/V^2}$ (ferromagnetic configuration).  This corresponds to an excited-state
phase transition~\cite{D_mondal_CT_2022}. When dissipation is introduced ($\gamma \neq 0$), the pitchfork structure collapses into a single stable branch, 
\begin{equation}
	\text{ FP-III:}\;
\begin{cases} z_1^*=z_2^*=-\sqrt{1-\dfrac{J^2}{V^2+\gamma^2}}\nonumber\\  \phi_i^*=\sin^{-1}(\gamma\sqrt{1-z_i^{*2}}/J)+\pi. 
\end{cases}
\end{equation}
This is shown in Fig.~\ref{Fig3}(a) with a single blue solid line for $V>V_c$. This single branch acts as a global attractor governing the long-time dynamics. The relaxation toward FP-III is illustrated in Fig.~\ref{Fig3}(c), which shows a classical trajectory of the normalized spin vector $\vec s_i(t)$ on the Bloch sphere converging to the attractor. Since $s_{iz}=z_i$, this convergence corresponds to the saturation of the population imbalance at a finite self-trapped value.

For $\gamma=0$, the fixed point FP-II also undergoes a pitchfork bifurcation at $V=V_c$. This transition breaks the left-right ($\mathbb{Z}_2$) symmetry and leads to an antisymmetric population imbalance, $z_1^*=-z_2^*\neq0$, in which the two species self-trap in opposite wells (antiferromagnetic configuration). These antisymmetric solutions persist as non-attracting fixed points when dissipation is introduced, 
\begin{eqnarray}
\text{FP-IV:} 
\;
\begin{cases} z_1^*=-z_2^*=\pm\sqrt{1-\dfrac{J^2}{V^2+\gamma^2}} \nonumber\\   \phi_i^*=-\sin^{-1}(\gamma\sqrt{1-z_i^{*2}}/J) ,
\end{cases}
\end{eqnarray}
as shown in Fig.~\ref{Fig3}(b). As a result, periodic orbits emerge in the vicinity of FP-IV, while other trajectories, depending on their initial conditions, may transiently explore their neighborhood before ultimately converging to the stable attractor FP-III at long times. This behavior is illustrated in Fig.~\ref{Fig3}(c), which shows representative classical trajectories  $\vec s_i(t)$ on the Bloch sphere.

Figure~\ref{Fig3}(d) shows the effect of dissipation on the self-trapping dynamics at the quantum level. The population-imbalance difference $\langle \hat{z}_{-} \rangle=(\langle \hat{z}_1\rangle-\langle \hat{z}_2\rangle)/2$
averaged over quantum trajectories ultimately decays to zero, consistent with the classical prediction that the long-time dynamics is governed by the stable attractor FP-III, where $z_1^{*}=z_2^{*}\ne0$. Interestingly, when the system is initialized in coherent states prepared near the non-attracting fixed points FP-IV, the dynamics remains confined to their vicinity for an extended transient time (blue curve) before relaxing to FP-III. In contrast, trajectories starting from coherent states centered at generic phase-space points (red curve) rapidly deviate from initial values and converge to the attractor. This behavior demonstrates that, despite the presence of dissipation, FP-IV still leaves a clear imprint on the transient dynamics when $V_c<V<\widetilde{V}_c$. 

For $V>\widetilde V_c$, the antisymmetric fixed points FP-IV becomes unstable. The boundary separating the regime where FP-IV can leave a transient dynamical imprint from the regime where it is unstable is marked by the dashed red line at $\widetilde V_c$ in Fig.~\ref{Fig1}(b).

\begin{figure*}[t]
\centering
\includegraphics[width=\linewidth]{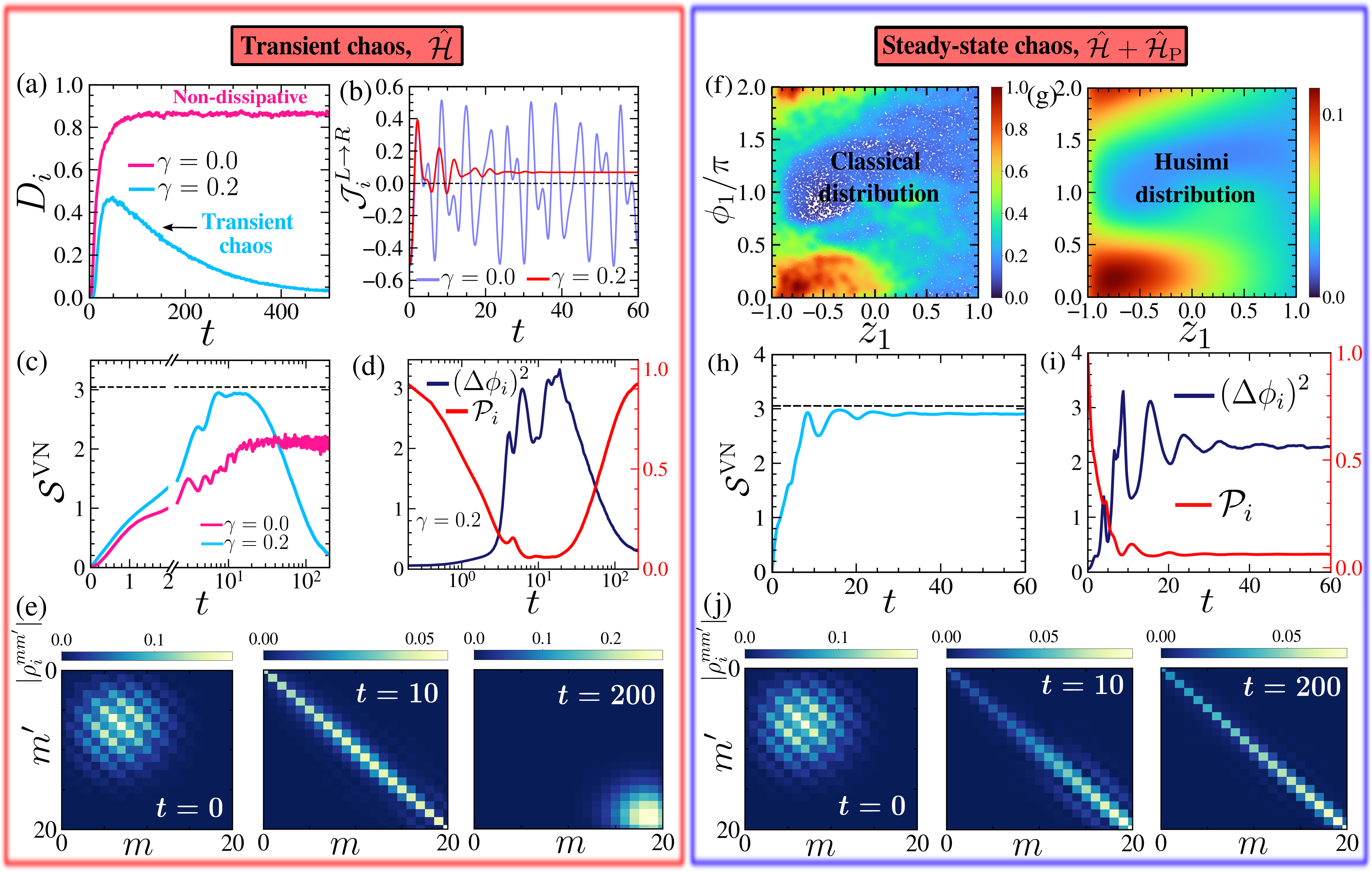}
\caption{{\bf Control of chaos and coherence via dissipation and a tilted potential.} 
Left panels (a)-(e): {\em Transient chaos and no tilted traps.} 
(a)-(b): Classical chaos in the absence of dissipation ($\gamma=0.0$, pink)  becomes transient chaos in its presence  ($\gamma=0.2$, blue) as shown with 
(a) decorrelator  and 
(b) atomic current of one species.   
(c)-(d): Corresponding quantum behavior. 
(c) Von Neumann entropy of one species for $\gamma=0$ and $0.2$; horizontal dashed line represents the maximum value for the open system $\mathcal{S}_{\rm max, open}^{\rm VN} 
 = \ln(2S+1) = \mathcal{S}_{\rm max, closed}^{\rm VN} +1/2$. The time is in the linear scale for $t\in[0,2]$ and in log scale for $t>2$. 
(d) Dynamics of purity $\mathcal{P}_i$ (right $y$-axis) and phase fluctuation $(\Delta\phi_i)^2$  (left $y$-axis) for $\gamma=0.2$. 
(e) Reduced density matrix of one species: Off-diagonal elements decay at short times ($t=10$), but coherence is recovered at long times ($t=200$). 
Right panels (f)-(j): {\em Steady-state chaos in presence of tilted traps, $\omega_z=0.5$.} All of these panels include dissipation, $\gamma=0.2$. 
%
(f) Long-time classical phase-space density on the chaotic attractor. (g) Husimi distribution $Q(z_1,\phi_1)$ of reduced density matrix $\hat{\rho}_1$ obtained from the long-time density matrix $\hat{\rho}$.  (h)-(i) Quantum dynamics for (h) the von Neumann entropy of one species,  (i) purity $\mathcal{P}_i$ (right y-axis), and phase fluctuation $(\Delta\phi_i)^2$ (left y-axis); saturating values indicate steady-state chaos.
(j) Reduced density matrix of one species: Coherence is no longer recovered at long times. 
All panels: $V=1.7$.  All quantum results are obtained for $S=10$, and the quantities in (c)-(e) and (h)-(j) are computed from $\hat{\rho}(t)$.
}
\label{Fig4}
\end{figure*}

\textbf{Transient chaos and coherence recovery:} Beyond signaling a phase transition, the critical interaction strength $V_c$ also marks the onset of chaotic dynamics in the isolated system. Introducing incoherent hopping qualitatively reshapes this behavior. Dissipation gives rise to a regime of \emph{transient} chaos~\cite{Deb_OADM_2026}, in which chaotic signatures persist only at intermediate times before the dynamics is ultimately drawn toward the stable attractor (FP-III). This relaxation is accompanied by a recovery of coherence at late times.

Figure~\ref{Fig4} illustrates the impact of transient chaos on both the classical [Figs.~\ref{Fig4}(a)–(b)] and quantum [Figs.~\ref{Fig4}(c)–(e)] dynamics. To diagnose classical chaos, Fig.~\ref{Fig4}(a) employs the decorrelator~\cite{Decorr1}
\begin{equation}
 D_i(t) = 1 - \langle \boldsymbol{s}_{i}^a . \boldsymbol{s}_{i}^b \rangle,   
\end{equation}
where $a$ and $b$ label two copies of the system evolved from slightly different initial conditions, and the average is taken over an ensemble of initial phase-space points. This quantity measures the growth and spatial spreading of initially localized perturbations and provides a classical analogue of the out-of-time-ordered correlator (OTOC), widely used to characterize quantum chaos. For $V>V_c$, Fig.~\ref{Fig4}(a) shows that in both the isolated and dissipative cases the decorrelator initially grows exponentially, with a rate analogous to a Lyapunov exponent. However, in contrast to the isolated system, dissipation causes $D_i(t)$ to decay at long times, approaching zero as the dynamics converges to the stable attractor.

The role of the attractor is also evident in the evolution of the atomic current $\mathcal{J}_i^{L\rightarrow R}$, which saturates to a nonzero value, as shown in Fig.~\ref{Fig4}(b). While oscillations persist for $\gamma=0$, they are rapidly suppressed by dissipation $\gamma\ne0$. 

Figure~\ref{Fig4}(c) displays the corresponding quantum behavior via the von Neumann entropy $\mathcal{S}^{\rm VN} = -{\rm Tr}(\hat{\rho}_i\ln\hat{\rho}_i)$ of the reduced density matrix $\hat{\rho}_i = {\rm Tr}_{\bar{i}}(\hat{\rho}(t))$ obtained by tracing out one of the two bosonic components from the full density matrix $\hat{\rho}(t)$. Similar to the classical dynamics in Fig.~\ref{Fig4}(a), in both the isolated and dissipative systems, $\mathcal{S}^{\rm VN}(t)$ initially grows linearly in time, indicating the rapid buildup of correlations and chaotic mixing.  Subsequently, the entropy saturates near the maximal value $\mathcal{S}_{\rm max}^{\rm VN}$, 
while in the presence of dissipation $\mathcal{S}^{\rm VN}(t)$ decreases and approaches zero as the system relaxes toward the attractor. 

This nonmonotonic entropy dynamics observed in the transient-chaos regime shows qualitative similarities to the Page-curve-like behavior originally discussed in the context of black-hole evaporation~\cite{Page_1993} and recently reported in interacting open quantum systems~\cite{Sarang_2025}. In those cases, entropy decreases due to relaxation toward a low-temperature steady state. In the present case, the entropy suppression is not caused by thermal relaxation, but by the emergence of a stable dynamical attractor that effectively constrains the accessible phase space and restores coherence at long times.

As dissipation suppresses chaotic dynamics, coherence is correspondingly restored. This is evidenced by the subsystem purity, $\mathcal{P}_i = {\rm Tr}(\hat{\rho}_i^2)$, shown in Fig.~\ref{Fig4}(d). The purity drops during the transient chaotic stage but recovers at long times, signaling coherence revival. Likewise, the phase fluctuations $(\Delta \phi_i)^2$, which are amplified during the chaotic evolution, decrease as the dynamics converges to the attractor, indicating that each condensate retains a high degree of phase coherence at long times. 

The coherence recovery is further illustrated in Fig.~\ref{Fig4}(e), which shows the time evolution of the matrix elements of $\hat{\rho}_i(t)$. While the off-diagonal elements are strongly suppressed at intermediate times (e.g., $t=10$), they partially reemerge at long times (e.g., $t=200$). Altogether, these results demonstrate that dissipation can be harnessed to mitigate the deleterious effects of chaos, restore coherence at late times, and tune the duration of the transient-chaotic window (see SI~\cite{SM}).

\textbf{Control of chaos with a tilted potential:} Additional control over the system's dynamical behavior can be achieved by introducing a tilt between the two wells. This tilt is described by the term
\[
\hat{\mathcal{H}}_{\rm P} = \frac{\omega_z}{2}\sum_i \bigl(\hat{n}_{i\rm L}-\hat{n}_{i\rm R}\bigr),
\]
which is added to the Hamiltonian in Eq.~(\ref{BJJ}). Physically, the tilt  breaks the left-right symmetry between the junction, changing the phase space structure.  For a finite range of the tilt amplitude $\omega_z$ (see SI~\cite{SM}), this perturbation destabilizes the attractor FP-III. As a result, the long-time relaxation responsible for coherence recovery is suppressed, and the system enters a qualitatively distinct dynamical regime characterized by persistent chaotic motion and referred to as steady-state chaos. 

The dynamical signatures of steady-state chaos are summarized in Figs.~\ref{Fig4}(f)–(j). Figure~\ref{Fig4}(f) shows the long-time classical phase-space density obtained by sampling trajectories from an ensemble of initial conditions in the presence of both dissipation and a finite tilt. Rather than collapsing onto a single attractor, the trajectories densely fill an extended region of phase space, revealing a chaotic attractor.

Quantum mechanically, the signature of the chaotic attractor can be obtained from the Husimi distribution
\begin{eqnarray}
Q(z_i,\phi_i) = \frac{1}{\pi}\bra{z,\phi}\hat{\rho}_i\ket{z,\phi}
\end{eqnarray}
of the reduced density matrix $\hat{\rho}_i$ of one of the bosonic species at long times. The Husimi function in Fig.\ref{Fig4}(g) closely mirrors the classical phase-space distribution, spreading over a large region instead of localizing around a fixed point or a regular orbit, thereby signaling steady-state chaos.

The persistence of chaos has profound consequences for coherence and information scrambling. As shown in Fig.~\ref{Fig4}(h), the von Neumann entropy of the reduced density matrix of one species grows rapidly and then saturates at a large value, in stark contrast to the decay observed in the transient-chaos regime. Consistently, Fig.~\ref{Fig4}(i) shows that the subsystem purity remains strongly suppressed at long times, while the phase fluctuations saturate at large values. These diagnostics indicate irreversible decoherence and sustained mixedness. 

This qualitative change is further highlighted by comparing the time evolution of the reduced density matrix elements in Fig.~\ref{Fig4}(j) with those in Fig.~\ref{Fig4}(e). In both transient [Fig.~\ref{Fig4}(e)] and steady-state [Fig.~\ref{Fig4}(j)] chaotic regimes, the off-diagonal elements are initially suppressed, reflecting rapid dephasing and information scrambling. However, only in the transient-chaos case do these coherences reemerge at long times. This contrast demonstrates that the duration of scrambling is not determined by interactions alone, but can be engineered through dissipative mechanisms. The ability to tune whether coherence is ultimately restored or permanently suppressed, by adjusting tilt and interaction strength, is a central result of our work.  

In the SI~\cite{SM}, we further analyze how the combined effect of interaction strength, dissipation, and tilt  magnitude can be used to tune not only the onset but also the duration of chaotic dynamics. Taken together, these results establish the open two-component BJJ as a versatile platform for exploring dissipative chaos, in which two qualitatively distinct regimes --  transient chaos with coherence recovery and steady-state chaos with irreversible decoherence -- can be selectively realized by adjusting experimentally accessible parameters.

From the perspective of Liouvillian spectral statistics, however, both transient and steady-state chaotic regimes exhibit the same Ginibre correlations. This reinforces the conclusions of Refs.~\cite{Lea_Santos_GHS,Deb_OADM_2026,Robnik_225}, which demonstrate a breakdown of the Grobe–Haake–Sommers conjecture~\cite{Haake_GHS}. The conjecture asserts that chaotic dynamics is characterized by Ginibre statistics in the Liouvillian spectrum, whereas regular dynamics gives rise to 2D Poisson statistics. As discussed in detail in the SI~\cite{SM}, our results show that Liouvillian spectral diagnostics predominantly probe short-time dynamical instability rather than the fate of the system at long-time.

\section{DISCUSSION}

Our work shows that dissipation can be harnessed to control not only the emergence of chaos but its lifetime, offering a new route to engineering dynamical phases and information flow in open quantum systems. In the two-component Bose-Josephson junction studied here, environment-induced incoherent hopping does not simply degrade coherence; rather, it reorganizes the phase space, retaining synchronized oscillations, rendering self-trapping stable, and enabling controlled access to qualitatively distinct chaotic regimes. In particular, dissipation allows quantum information to be temporarily scrambled and von Neumann entropy to grow, while still enabling the recovery of coherence at long times.

\section{MATERIALS AND METHODS}

\textbf{Classical analysis:} Starting from the master equation, the time evolution of the expectation value of any operator $\hat A$ is given by $\frac{d}{dt}\langle \hat A\rangle = \mathrm{Tr}(\dot{\rho}\,\hat A)$.
The classical equations of motion (EOM) for the conjugate variables $\{\phi_i, z_i\}$, presented in the main text [Eq.~\eqref{dissipative_EOM}], are obtained by applying this relation within the decoupling (mean-field) approximation $\langle AB\rangle=\langle A\rangle\langle B\rangle$. This approximation becomes exact in the limit of large spin magnitude $S\rightarrow\infty$~\cite{Souza2023,Exactness_1,Exactness_2}.

The fixed points $\{z_i^*,\phi_i^*\}$ corresponding to steady states are determined by setting $\dot z_i=\dot\phi_i=0$ in the EOM. Their stability is analyzed via linear stability analysis~\cite{SM}, by examining the eigenvalues $\lambda$ governing the time evolution of small fluctuations about the fixed points. If all eigenvalues satisfy $\mathrm{Re}(\lambda)<0$, the fixed point is a stable attractor, and trajectories within its basin of attraction converge toward it. If all eigenvalues have vanishing real parts, $\mathrm{Re}(\lambda)=0$, the fixed point is a non-attracting center, around which small-amplitude fluctuations lead to oscillatory motion with frequencies $|\mathrm{Im}(\lambda)|$. The presence of any eigenvalue with $\mathrm{Re}(\lambda)>0$ signals an instability.

Chaotic dynamics is characterized by an enhanced sensitivity to initial conditions, whereby small deviations grow exponentially in time as $\sim e^{\Lambda_l t}$. The maximal Lyapunov exponent $\Lambda_l$, defined as the largest positive growth rate and evaluated numerically at long times, provides a quantitative measure of fully developed steady-state chaos~\cite{Strogatz}. In addition, chaos can be diagnosed through the time evolution of the decorrelator $D(t)$~\cite{Decorr1}, which captures the divergence of nearby trajectories. 
\\

\textbf{Quantum trajectories:}
To study the open quantum dynamics of the dissipative two-component BJJ, we employ the stochastic wave-function (quantum trajectory) method~\cite{Daley_2014}. This approach provides an efficient description of the Lindblad dynamics in terms of an ensemble of pure-state trajectories.

For the quantum dynamics, we take the initial state to be a product state
$ \ket{\psi_c}=\ket{\theta_1,\phi_1}\otimes\ket{\theta_2,\phi_2}$,
where $\ket{\theta_i,\phi_i}$ are spin-coherent states associated with the two collective spins, defined as
\begin{equation}
\ket{\theta_i,\phi_i}
=
\cos^{2S}\!\left(\frac{\theta_i}{2}\right)
\exp\!\left[
e^{-i\phi_i}\tan\!\left(\frac{\theta_i}{2}\right)\hat{S}_{i-}
\right]
\ket{S,S}.
\end{equation}
These states provide a semiclassical representation of the phase-space point
$\{z_i=\cos\theta_i,\phi_i\}$.
In the numerical simulations, we consider a finite spin magnitude $S$.

Within the stochastic wave-function framework, the state is evolved between quantum jumps according to the non-Hermitian effective Hamiltonian
\begin{equation}
\hat{\mathcal{H}}_{\rm NH}
=
\hat{\mathcal{H}}
-
\frac{i}{2}\sum_i
\hat{\mathcal{O}}_i^{\dagger}\hat{\mathcal{O}}_i,
\label{Non_Hermitian_hamiltonian}
\end{equation}
where $\hat{\mathcal{O}}_i$ are the jump operators defined in Eq.~\eqref{Master_equation}. At each time step $dt$, a quantum jump associated with operator $\hat{\mathcal{O}}_i$ occurs with probability
$\delta P_i = dt\,\langle \psi(t)|\hat{\mathcal{O}}_i^{\dagger}\hat{\mathcal{O}}_i|\psi(t)\rangle$.
When a jump occurs, the wave function is updated and normalized as
\begin{equation}
\ket{\psi(t+dt)} =
\frac{\hat{\mathcal{O}}_i\ket{\psi(t)}}
{\sqrt{\bra{\psi(t)}\hat{\mathcal{O}}_i^{\dagger}\hat{\mathcal{O}}_i\ket{\psi(t)}}}.
\end{equation}
Between jumps, the state continues to evolve under $\hat{\mathcal{H}}_{\rm NH}$.
This procedure is repeated to generate individual quantum trajectories.

The open-system dynamics is obtained by averaging over an ensemble of stochastic wave-function trajectories. The system density matrix is reconstructed as
\begin{eqnarray}
\hat{\rho}(t) = \frac{1}{N_{\rm traj}}\sum_{j=1}^{N_{\rm traj}} \ket{\psi_j(t)}\bra{\psi_j(t)},
\end{eqnarray}
where $N_{\rm traj}$ denotes the number of trajectories and $j$ labels each realization. In the limit of a sufficiently large ensemble, $N_{\rm traj} \gg 1$, this procedure converges to the solution of the Lindblad master equation.

Physical observables are computed from the ensemble-averaged density matrix $\hat{\rho}(t)$. To analyze subsystem properties, we evaluate the reduced density matrix
\[
\hat{\rho}_i(t)=\mathrm{Tr}_{\bar{i}}\!\left[\hat{\rho}(t)\right],
\]
obtained by tracing out the complementary component. From $\hat{\rho}_i(t)$ we compute quantities such as the von Neumann entropy, purity, and phase fluctuations, as discussed in the main text.
\\

\textbf{Phase fluctuations of the BJJ:}
Phase coherence is an important property of a Josephson junction and can be degraded both by enhanced quantum fluctuations and by the onset of chaotic dynamics.
To quantify phase fluctuations, we adopt a quantum description of the relative phase based on a discrete phase representation~\cite{Oberthalar_2}.
The phase eigenstates are defined as
\begin{equation}
\ket{\phi_m}
=
\frac{1}{\sqrt{2S+1}}
\sum_{n=-S}^{S}
e^{i n \phi_m}\ket{n},
\end{equation}
where
\(
\phi_m = -\pi + \frac{2\pi m}{2S+1}
\),
with integer $m\in\{0,1,\ldots,2S\}$ and $\phi_m\in[-\pi,\pi]$.
Here, $\ket{n}$ denotes the eigenstates of the population-imbalance operator.

The phase probability distribution associated with the reduced density matrix $\hat{\rho}_i$ of one species is given by
\[
p(\phi_m)
=
\bra{\phi_m}\hat{\rho}_i\ket{\phi_m},
\qquad
\sum_m p(\phi_m)=1.
\]
For each quantum trajectory, the phase fluctuation is computed from this distribution as
\begin{equation}
(\Delta\phi)^2
=
\sum_m
\bigl(\phi_m-\langle\phi\rangle\bigr)^2
\,p(\phi_m),
\end{equation}
where
\(
\langle\phi\rangle=\sum_m \phi_m\,p(\phi_m)
\)
denotes the mean phase.
By repeating this procedure for an ensemble of quantum trajectories, we obtain the time evolution of the phase fluctuations discussed in the main text.
\\

\textbf{Acknowledgements:}
LFS was supported by the Research Corporation Cottrell SEED award CS-SEED-2025-003.

\bibliography{bibliography.bib}

\end{document}


\begin{center}

{\large \bf Supplemental Information: \\Dissipation as a Resource: Synchronization, Coherence Recovery, and Chaos Control}\\

\vspace{0.6cm}

Debabrata Mondal$^1$, Lea F. Santos$^2$, and S. Sinha$^1$\\

$^1${\it Indian Institute of Science Education and Research-Kolkata,
Mohanpur, Nadia-741246, India}

$^2${\it Department of Physics, University of Connecticut, Storrs, Connecticut 06269, USA}

\end{center}


\renewcommand{\theequation}{S\arabic{equation}}
\renewcommand{\thefigure}{S\arabic{figure}}

This supplemental information provides additional figures and analyses that support the discussions in the main text. 

\section*{Supplementary note 1: \\Derivation of the dissipative coupled-top model and its classical dynamics}
\label{Section1}

The two-component bosonic Josephson junction (BJJ) discussed in the main text can be mapped onto a system of two coupled large spins, referred to as the coupled-top (CT) model \cite{D_mondal_CT_2022}. Within the Schwinger boson representation, 
\[
\hat{S}_{i-}= \hat{a}_{i{\rm R}}^{\dagger}\hat{a}_{i{\rm L}} \qquad \hat{S}_{iz}=(\hat{n}_{i{\rm L}}-\hat{n}_{i{\rm R}})/2,
\] 
the BJJ Hamiltonian [Eq.(1) of the main text] maps to the CT Hamiltonian
\begin{eqnarray}
	\hat{\mathcal{H}}=-J\hat{S}_{1x}-J\hat{S}_{2x}+\frac{V}{S}\hat{S}_{1z}\hat{S}_{2z},
\end{eqnarray}
where $i=1,2$ labels the two spins associated with the two bosonic species, $\hat S_{ia}$ ($a=x,y,z$) are spin components, and both spins have magnitude $S=N/2$.
 
Similarly, the dissipator describing the incoherent tunneling process can be written in the spin language using the jump operators
\begin{equation}
\hat{\mathcal{O}}_i=\sqrt{\frac{\gamma}{S}}\,\hat S_{i-}.
\end{equation} 
As a result, the master equation given in Eq.(2) of the main text describes the dissipative CT model. Since this model possesses a well-defined classical limit for $S\rightarrow \infty$, it facilitates a direct quantum-classical correspondence \cite{D_mondal_CT_2022}. In the classical limit, each spin is represented by a unit vector $\vec{S}_i/S \equiv\vec{s}_i = \{\sin(\theta_i)\cos(\phi_i),\sin(\theta_i)\sin(\phi_i), \cos(\theta_i)\}$ on the Bloch sphere, where the canonically conjugate variables are the angle $\phi_i$ and $z_i=\cos\theta_i$.  The classical equations of motion quoted in Eq.~(5) of the main text can be derived from the master equation in the classical limit (see the Methods section of the main text).

{\bf Fixed points and linear stability.}
The non-equilibrium phases (steady states) of the open CT model are characterized by the fixed points $\mathbf{X}^*=\{\phi_1^*,z_1^*,\phi_2^*,z_2^*\}$ of the classical equations of motion (EOM) [Eq.(5) of the main text], obtained by setting $\dot{\mathbf{X}}=0$. This yields the four fixed points FP-I $-$ FP-IV discussed in the main text. 

The stability of each fixed point is determined by linearizing the EOM around $\mathbf{X}^*$. A small deviation $\delta \mathbf{X} (t)$ evolves as $\delta \mathbf{X}(t)=\delta \mathbf{X} e^{\lambda t}$, where $\lambda$ are the (generally complex) eigenvalues of the stability matrix~\cite{Strogatz}. A fixed point is a stable attractor if ${\rm Re}(\lambda)<0$ for all eigenvalues; in that case, trajectories converge to it. In the open CT model, FP-I, FP-II, and FP-IV are centers with ${\rm Re}(\lambda)=0$, supporting periodic orbits analogous to those in Hamiltonian dynamics, while FP-III is an attractor, so nearby trajectories converge to it. The frequency of small-amplitude oscillations about a center is given by $\omega={\rm Im}(\lambda)$~\cite{Strogatz}, as reported in Eq.~(8) of the main text for FP-I and FP-II. 

{\bf Exchange symmetry and the symmetric dynamical class.}
We next discuss the emergence of periodic orbits and  synchronized oscillation between the two spins. The open CT model and the corresponding classical EOM [Eq.(5) of the main text] are invariant under the exchange of two spins $(S_1 \leftrightarrow S_2)$. 
To understand the effect of this exchange symmetry on the dynamics, we rewrite the EOM in terms of the new classical variables 
\[
z_\pm = (z_1 \pm z_2)/2 \qquad \phi_\pm = (\phi_1 \pm \phi_2)/2,
\]
which lead to the equations of motion,
\begin{subequations}
\begin{align}
\dot z_+ &= -\frac{J}{2}\Big[
\sqrt{1-(z_+ + z_-)^2}\,\sin(\phi_+ + \phi_-)
+\sqrt{1-(z_+ - z_-)^2}\,\sin(\phi_+ - \phi_-)
\Big] -\frac{\gamma}{2}\Big[2-(z_+ + z_-)^2-(z_+ - z_-)^2\Big], \label{eq:zplus} \\[1ex]
\dot\phi_+ &= \frac{J}{2}\Big[
\frac{(z_+ + z_-)\cos(\phi_+ + \phi_-)}{\sqrt{1-(z_+ + z_-)^2}}
+\frac{(z_+ - z_-)\cos(\phi_+ - \phi_-)}{\sqrt{1-(z_+ - z_-)^2}}
\Big]
+ V z_+, \label{eq:phiplus} \\[1ex]
\dot z_- &= -\frac{J}{2}\Big[
\sqrt{1-(z_+ + z_-)^2}\,\sin(\phi_+ + \phi_-)
-\sqrt{1-(z_+ - z_-)^2}\,\sin(\phi_+ - \phi_-)
\Big] +\frac{\gamma}{2}\Big[(z_+ + z_-)^2-(z_+ - z_-)^2\Big], \label{eq:zminus} \\[1ex]
\dot\phi_- &= \frac{J}{2}\Big[
\frac{(z_+ + z_-)\cos(\phi_+ + \phi_-)}{\sqrt{1-(z_+ + z_-)^2}}
-\frac{(z_+ - z_-)\cos(\phi_+ - \phi_-)}{\sqrt{1-(z_+ - z_-)^2}}
\Big]
- V z_-. \label{eq:phiminus}
\end{align}
\label{EOM_new_variables}
\end{subequations}
It is evident from Eqs.\eqref{EOM_new_variables}(c,d) that the condition $\{z_-=0, \,\phi_-=0\}$ is a solution of the EOM, describing a special class of constrained dynamics that we refer to as the {\it symmetric dynamical class}, since the two spins are completely aligned with the condition $s_{-x}=s_{-y}=s_{-z}=0$.

Consequently, the effective dynamics in the reduced phase space of the remaining variables $\{z_+,\phi_+\}$ are described by the following EOM,
\begin{eqnarray}		
\dot{z}_+=-J\sqrt{1-z_+^2}\sin{\phi_+}-\gamma(1-z_+^2),\qquad\dot{\phi}_+=\frac{Jz_+\cos{\phi_+}}{\sqrt{1-z_+^2}}+V z_{+}.
\label{LMG_EOM}
\end{eqnarray}
Notably, the evolution of $\{z_+,\phi_+\}$ governed by the above equations corresponds to the dissipative dynamics of the Lipkin-Meshkov-Glick (LMG) model with antiferromagnetic interaction \cite{LMG}. Interestingly, this system can exhibit regular periodic orbits even in the presence of dissipation,  due to the emergence of a conserved quantity, 
\begin{eqnarray}
\mathcal{R} = 2{\rm Re}\left[(-i\gamma-V)\ln\left(i\sqrt{1-z_+^2}\exp{(-i\phi_+)}+\frac{J}{\gamma-iV}\right)\right].
\label{Conserved_quantity}
\end{eqnarray}
Therefore, for initial conditions in the symmetric dynamical class, $\{z_-=0, \,\phi_-=0\}$, the open CT model exhibits synchronized oscillatory motion in which both spins remain aligned. This behavior is akin to boundary time crystals \cite{Boundary_time_crystal}.

To assess the stability of this synchronized motion, we consider small perturbations that violate the constraint, $z_-\neq 0$ and/or $\phi_-\neq 0$. For $V<V_c$, we find that $\mathcal{R}$ remains quasi-conserved, displaying only small fluctuations about a constant value. As a consequence, quasi-periodic orbits persist in the full phase space, as discussed in the main text. 

In the next section, we investigate the stability of the synchronized periodic dynamics under perturbations that violate the symmetric dynamical class, considering both $V<V_c$ and $V>V_c$. In the latter case, $\mathcal{R}$ is no longer quasi-conserved, because the onset of transient chaos destroys the synchronized motion.

\section*{Supplementary note 2: \\Semiclassical truncated Wigner approximation}
\label{Section2}

As noted before, oscillatory dynamics, including time crystals, is most pronounced for large spins, $S\gg 1$. In the present setup of the BJJ, the number of atoms used in the experiments is typically large ($N\sim 10^3-10^4$), and quantum fluctuations are suppressed because the effective Planck constant scales as $\hbar/N$. 
For sufficiently large $N$, the dynamics of the open BJJ can be analyzed semiclassically using the truncated Wigner approximation (TWA) \cite{TWA_1,TWA_2,TWA_Spin}.  Within TWA, quantum fluctuations are incorporated by sampling the initial conditions from a Gaussian distribution of width $\sim 1/S$ around a chosen classical phase-space point and evolving each sample  following the stochastic equations,
\begin{subequations}
\begin{eqnarray}
\dot{s}_{ix} &=& -V s_{iy}s_{\bar{i}z}+\gamma s_{ix}s_{iz} + \sqrt{\frac{\gamma}{S}}\,\,s_{iz}\,\xi_{1i}\\
\dot{s}_{iy} &=& Js_{iz}+V s_{ix}s_{\bar{i}z}+\gamma s_{iy}s_{iz}- \sqrt{\frac{\gamma}{S}}\,\,s_{iz}\,\xi_{2i}\\
\dot{s}_{iz} &=& -Js_{iy}-\gamma(s_{ix}^2+s_{iy}^2)- \sqrt{\frac{\gamma}{S}}\,\,s_{ix}\,\xi_{1i} +\sqrt{\frac{\gamma}{S}}\,\,s_{iy}\,\xi_{2i},
\end{eqnarray}
\end{subequations}
where $\bar{i}$ denotes the complementary species ($\bar{1}=2$, $\bar{2}=1$) and the real Gaussian noise satisfies \cite{TWA_Spin} 
\[
\langle \xi^*_{ji}(t)\xi_{j'i'}(t')\rangle=
\delta_{jj'}\delta_{ii'}\delta(t-t') .
\]
Each initial state is evolved independently and the ensemble average of an observable is computed as
\[
\langle O\rangle_{\scriptscriptstyle\text{TWA}} (t) = \sum_j O_j(t)/{\mathcal{N}},
\]
where $\mathcal{N}$ is the number of trajectories with index $j$. 

{\bf Synchronization and its breakdown.}
Next, we investigate the synchronized oscillation between the two spins corresponding to the dynamical class $\{s_{-x}=s_{-y}=s_{-z}=0\}$ using TWA for a finite spin magnitude $S$. For this purpose, we consider a product coherent state of two aligned spins and sample initial conditions from the corresponding Husimi distribution. Note that this sampling also includes fluctuations that violate the symmetric constraint.

\begin{figure}[h]
\centering
\includegraphics[width=0.65\linewidth]{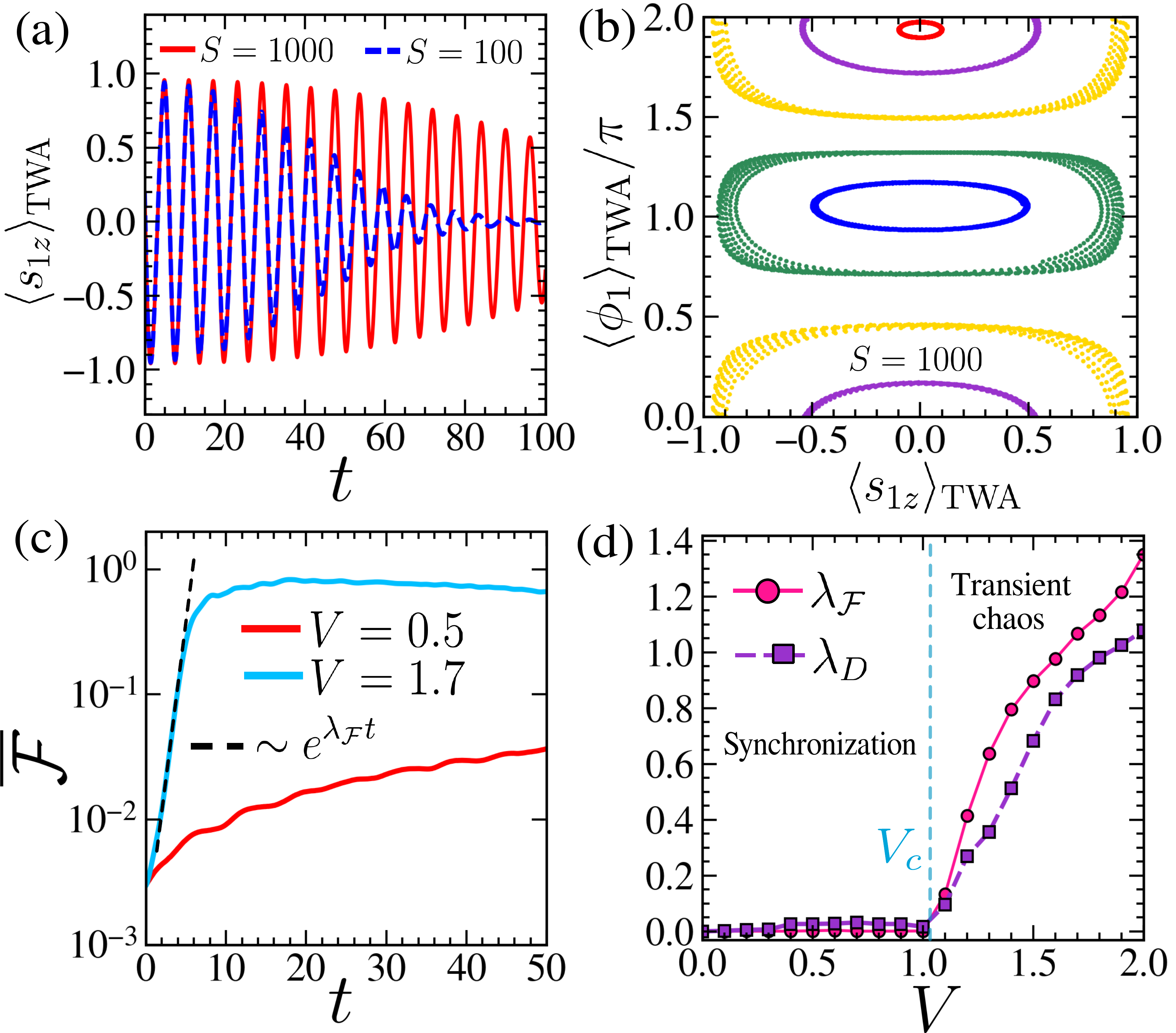}
\caption{
{\bf Synchronized oscillatory dynamics within the truncated Wigner approximation}. (a) Time evolution of $\langle s_{1z}\rangle_{\rm TWA}$ for spin magnitudes $S=100,1000$. (b) Trajectories on the plane of conjugate variables $\langle s_{1z}\rangle_{\rm TWA}$ and $\langle \phi_1\rangle_{\rm TWA}$ for $S=1000$, starting from initial conditions in the symmetric dynamical class at $V=0.5<V_c$. (c) Dynamics of the average fluctuation measure $\overline{\mathcal{F}}$ [Eq.\eqref{Fluctuation}] for coupling strengths $V=0.5,1.7$. (d) Dependence of the exponential growth rates $\lambda_{\mathcal{F}}$ and  $\lambda_{D}$ of the fluctuation and the decorrelator on the interaction strength $V$, exhibiting a crossover from synchronized motion to transient chaos at $V=V_c$. The spin magnitude for panels (c,d) is $S=1000$. The dissipation strength for all panels is $\gamma=0.2$. Here and in rest of the figures, we choose $J=1$.
}
\label{Supp_Fig1}
\end{figure}

We find that the semiclassical TWA dynamics agrees well with the full quantum dynamics for moderate $S$. 
In the synchronized regime $V<V_c$, the oscillatory behavior is visible in the time evolution of $\langle s_{1z}\rangle_{\mathrm{TWA}}$, as shown in Fig.~\ref{Supp_Fig1}(a). 
As evident from Fig.\ref{Supp_Fig1}(a), the oscillations exhibit a slow decay, but the associated time scale grows with increasing $S$. The corresponding dynamics in the conjugate plane $\bigl(\langle s_{1z}\rangle_{\mathrm{TWA}},\langle \phi_1\rangle_{\mathrm{TWA}}\bigr)$ is shown in Fig.~\ref{Supp_Fig1}(b), where the trajectories form nearly periodic orbits with a weak diffusive broadening due to quantum fluctuations. These orbits look qualitatively similar to those obtained from the classical and quantum trajectory methods shown in Fig.2(d,e) of the main text.

When the two spins are synchronized, the differences of their components satisfy $\langle s_{-a}\rangle_{\text{\tiny TWA}}\equiv(\langle s_{1a}\rangle_{\text{\tiny TWA}}-\langle s_{2a}\rangle_{\text{\tiny TWA}})/2 \approx 0 \,\,(a=x,y,z)$  and the corresponding fluctuations $(\Delta s_{-a})^2=\langle s_{-a}^2\rangle_{\text{\tiny TWA}}-\langle s_{-a}\rangle_{\text{\tiny TWA}}^2$ remain small.
To analyze how these fluctuations grow, we study the dynamics of
\begin{eqnarray}
\mathcal{F}=(\Delta s_{-x})^2+(\Delta s_{-y})^2+(\Delta s_{-z})^2
\label{Fluctuation}
\end{eqnarray}
averaged over random initial phase space points chosen uniformly from the symmetric dynamical class, denoted as $\overline{\mathcal{F}}$.  As shown in Fig.\ref{Supp_Fig1}(c), $\overline{\mathcal{F}}$ grows slowly for $V<V_c(\gamma)$ exhibiting synchronization, whereas for $V>V_c(\gamma)$, it grows exponentially with rate $\lambda_{\mathcal{F}}$, indicating that transient chaos destroys synchronization.

To quantify how the onset of chaos hinders synchronization, we also compute the decorrelator $D_i$ for the $i$th spin (defined in the main text) using TWA and averaging over initial conditions restricted to the symmetric dynamical class.
Similar to  $\overline{\mathcal{F}}$,  the decorrelator exhibits exponential growth with rate $\lambda_D$ for $V>V_c$, while it grows slowly in the synchronized regime,  implying a regular oscillatory dynamics.
The dependence of growth rate of both fluctuation $\mathcal{F}$ and decorrelator $D_i$ on $V$ is shown in Fig.~\ref{Supp_Fig1}(d), highlighting the crossover from synchronized oscillations to transient chaos at the critical coupling strength $V_c$. This analysis also indicates that the decorrelator can be used as a detection tool for synchronization. Note that in the transient-chaotic regime, both the fluctuation and decorrelator eventually decay at long times due to convergence to the stable attractor FP-III (see Fig.~4(a) of the main text).

{\bf Dissipative phase transition.}
As discussed in the main text, a dissipative phase transition occurs at the critical interaction strength $V_c$, above which two distinct classes of self-trapped states emerge, one of which (FP-III) forms an attractor and the two branches of FP-IV correspond to the centers. 
The asymptotic dynamics is mainly governed by the attractor FP-III, since the regular regions in the phase space surrounding the centers FP-IV are comparatively small relative to the basin of attraction of FP-III. Consequently, the motion in the vicinity of FP-IV can be destabilized by fluctuations, and the trajectories eventually converge to FP-III.

\begin{figure}[t]
\centering
\includegraphics[width=0.5\linewidth]{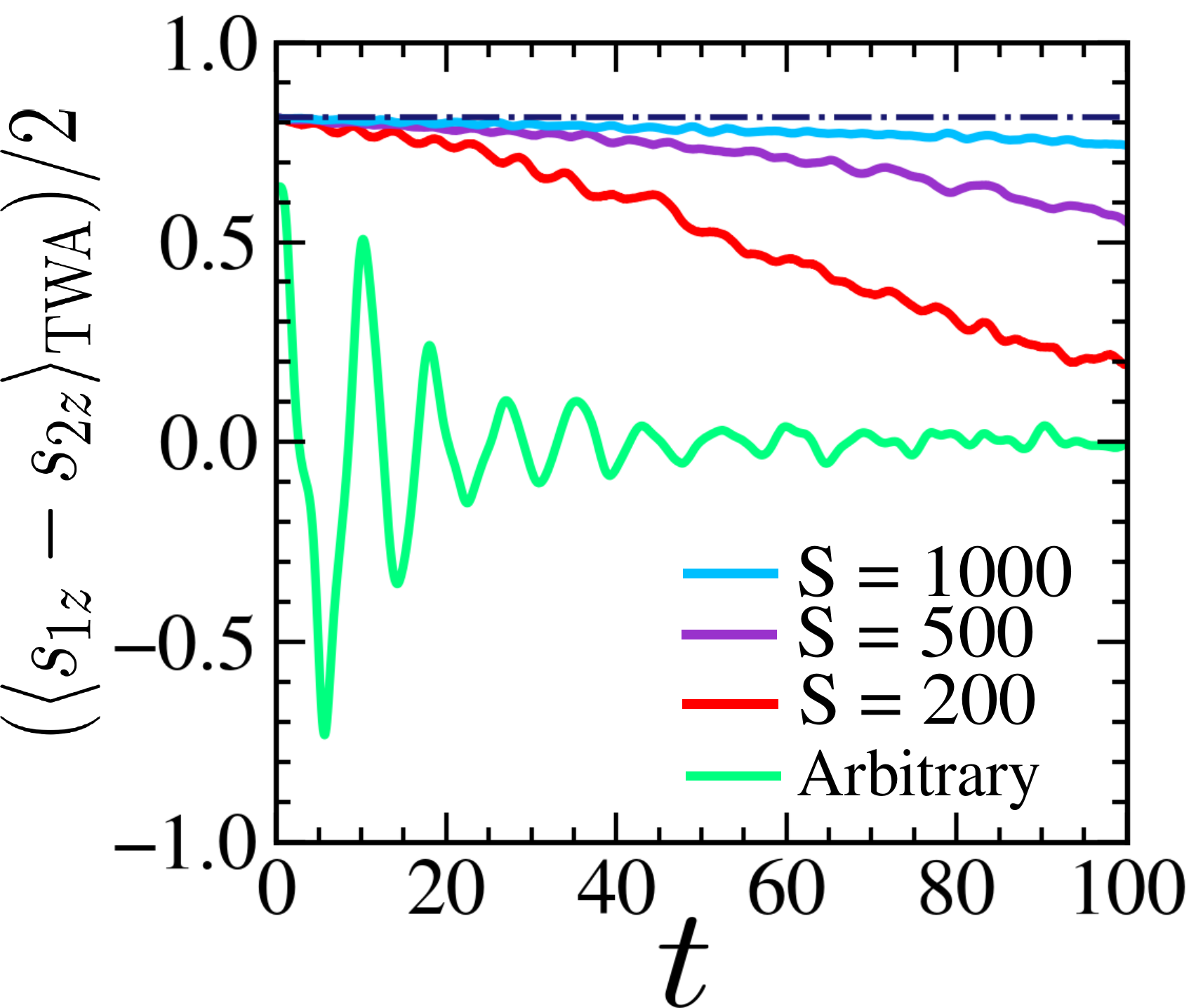}
\caption{
{\bf Signature of the self-trapped state due to the center FP-IV using the truncated Wigner approximation}. Dynamics of $\langle s_{-z}\rangle_{\text{\tiny TWA}}\equiv(\langle s_{1z}\rangle_{\text{\tiny TWA}}-\langle s_{2z}\rangle_{\text{\tiny TWA}})/2$ starting from FP-IV for different spin magnitudes $S$ and for an arbitrary initial state chosen from the chaotic region of phase space for $S=1000$. Parameters:  $\gamma=0.2$ and $V=1.7$.
}
\label{Supp_Fig2}
\end{figure}

Here, we investigate the signature of this non-attracting self-trapped state (FP-IV) semiclassically, within TWA. Specifically, we analyze the time evolution of $\langle s_{-z}\rangle_{\text{\tiny TWA}}$,
starting from the centers of FP-IV [see main text]. 
As shown in Fig.\ref{Supp_Fig2}, $\langle s_{-z}\rangle_{\mathrm{TWA}}$ slowly deviates from the FP-IV and asymptotically  approaches zero at long times, $\langle s_{-z}\rangle_{\text{\tiny TWA}}(t\rightarrow \infty)\rightarrow0$, which corresponds to the attractor FP-III. 
In contrast, starting from any arbitrary initial state from the chaotic region of the phase space, the dynamics converges to the attractor much more rapidly  (Fig.\ref{Supp_Fig2}).
The time scale over which the dynamics remains close to FP-IV increases with the spin magnitude $S$, indicating that FP-IV becomes stable in the classical limit $S\rightarrow\infty$.
%

\begin{figure}[h]
	\centering
	\includegraphics[width=0.45\linewidth]{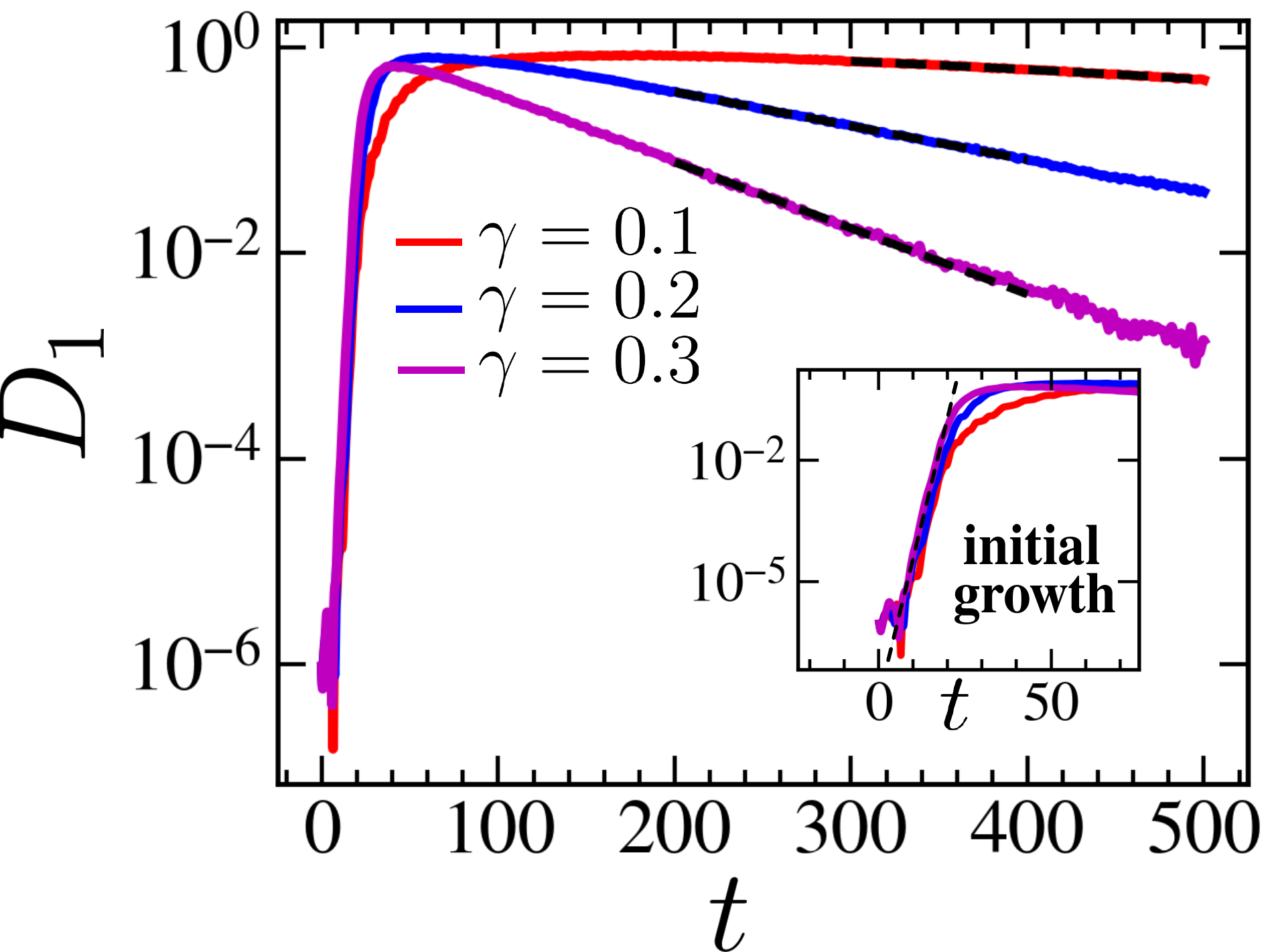}
	\caption{ {\bf Dependence of  transient chaos on the strength of dissipation}. Dynamics of the decorrelator $D_1$ obtained via TWA for different dissipation strengths $\gamma=0.1,0.2,0.3$ and coupling strength $V=1.7$. The inset highlights the early-time exponential growth. The initial states are sampled from a Gaussian distribution (for $S=1000$) from the chaotic region of the phase space.
	}
	\label{Supp_Fig1A}
\end{figure}

{\bf Dependence of transient-chaos suppression on $\gamma$:}
In the absence of dissipation ($\gamma=0$) and $V>V_c$, chaos is present in the corresponding Hamiltonian system. 
When dissipation is introduced, chaos is suppressed at long times due to the emergence of the stable attractor FP-III, giving rise to the transient chaos discussed in the main text.
 The rate of chaos suppression increases with $\gamma$ as the attractor FP-III becomes stronger. The degree of chaos can be captured from the dynamics of the decorrelator $D_i$ shown in Fig.\ref{Supp_Fig1A}. The early-time exponential growth rate of $D_i$ remains approximately  unchanged as the dissipation strength increases (inset), reflecting the underlying Hamiltonian-like instability at short times. However, at later times the decorrelator decays because of the attractor and the corresponding rate increases with $\gamma$.

\section*{Supplementary note 3: \\Oscillatory phase of the open BJJ}
\begin{figure}[b]
\centering
\includegraphics[width=0.8\linewidth]{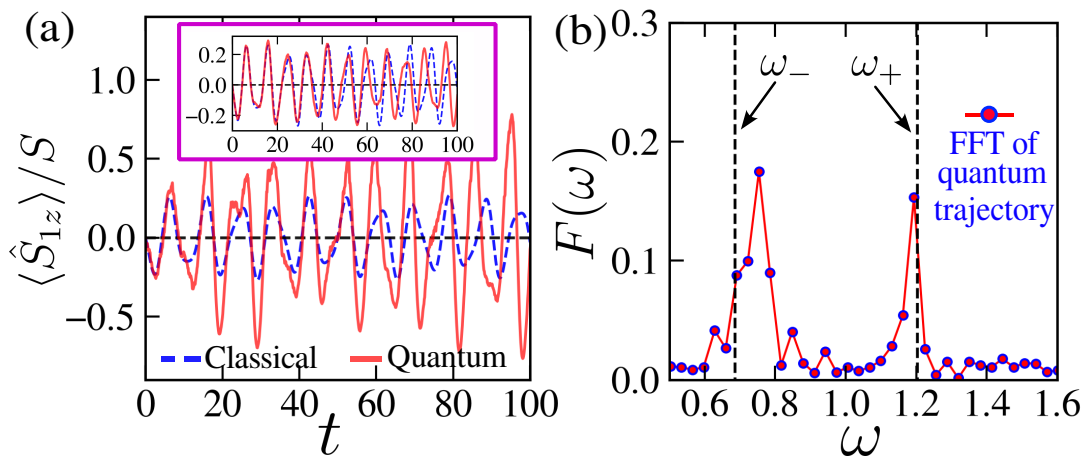}
\caption{
{\bf Time quasi-crystal in the open BJJ.} (a) Dynamics of the spin component (population imbalance) $\langle \hat{S}_{1z}\rangle$ obtained from a single quantum trajectory (main panel) and averaged over trajectories (inset) starting from an initial coherent state near FP-I. The blue dashed line represents the corresponding classical quasi-periodic trajectory starting from the same initial condition. (b) Fourier spectrum $F(\omega)$ of the single quantum trajectory in (a) reveals the frequencies $\omega_{\pm}$ [Eq.(8) of the main text] of the oscillation.  Parameters $S=50$, $V=0.5$, $\gamma=0.2$.
}
\label{Supp_Fig2A}
\end{figure}

In the main text, we discussed the time crystalline behavior of the condensates in the oscillatory regime $V<V_c$. Starting from identical initial configurations for both species, the condensates exhibit synchronized periodic oscillations. Here, we show that even when the initial states of the two condensate species are not exactly identical, oscillations persist but become quasi-periodic, with a small number of characteristic frequencies, thereby constituting a time quasi-crystal \cite{He2025PRX}. To capture the signatures of this unique time crystal, we consider an initial coherent state close to the classical fixed points FP-I and FP-II. As shown in Fig.~\ref{Supp_Fig2A}(a), the dynamics of the population imbalance (equivalently the spin component $\langle \hat{S}_{iz}\rangle$) obtained from individual quantum trajectories using the stochastic wave-function method (see Methods) exhibits quasi-periodic oscillations.
The Fourier spectrum of a single trajectory, shown in Fig.\ref{Supp_Fig2A}(b), reveals two characteristic frequencies  $\omega_{\pm}$ [Eq.(8) of the main text] corresponding to the small-amplitude oscillations around the fixed points FP-I and FP-II.
These results exhibit the usefulness of the quantum trajectory method, which can unveil coherent and oscillatory dynamics even for a small number of atoms $N$.

\begin{figure}[t]
\centering
\includegraphics[width=0.7\linewidth]{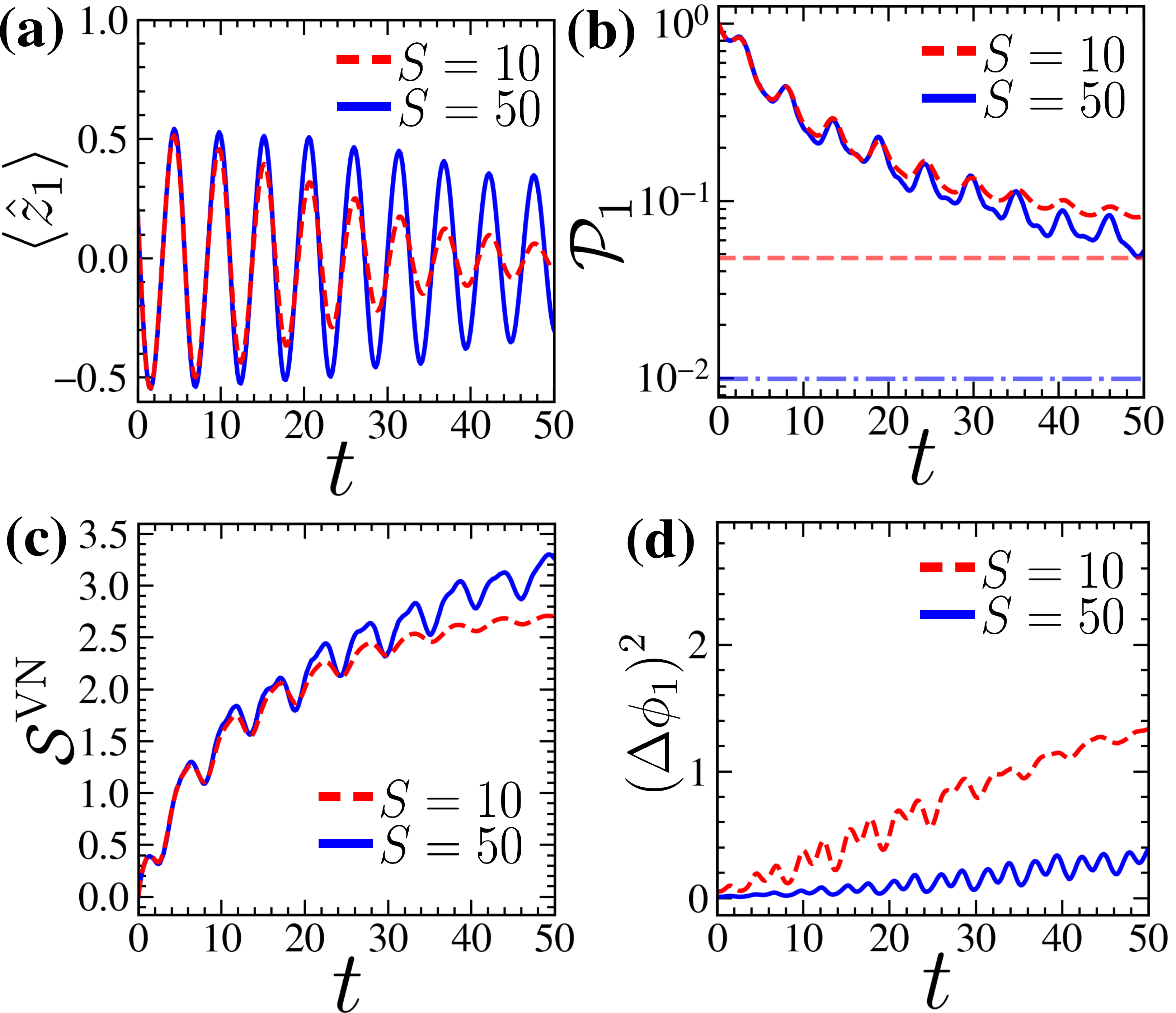}
\caption{
{\bf Coherence properties of the quantum states in the oscillatory phase.}  
Dynamics of (a) average population imbalance $\langle\hat{z}_1\rangle$, (b) purity $\mathcal{P}_1$, (c) von Neumann entropy $\mathcal{S}^{\rm VN}$, and (d) phase fluctuation $(\Delta \phi_1)^2$ for spin magnitudes $S=10,50$. The horizontal red dashed (blue dashed dotted) line indicates the maximally mixed purity $1/d$ with $d=2S+1$. Parameters: $V=0.5$, $\gamma=0.2$.
}
\label{Supp_Fig4}
\end{figure}

{\bf Coherence in the oscillatory regime:}
We analyze the coherent nature of the oscillatory dynamics for $V<V_c$  using the density matrix obtained from the stochastic wave-function method (see the main text). 
For finite spin magnitude $S$, the oscillation amplitude of the averaged population imbalance $\langle \hat{z}_i\rangle$ decays in time, but the corresponding time scale increases with increasing $S$, as shown in Fig.\ref{Supp_Fig4}(a).

To quantify coherence, we compute the reduced density matrix of one spin, 
\[
\hat{\rho}_i = {\rm Tr}_{\bar{i}}(\hat{\rho}(t)) \qquad \text{where} \qquad \hat{\rho}(t)=\frac{1}{N_{\rm traj}}\sum_{j=1}^{N_{\rm traj}}\ket{\psi_j}\bra{\psi_j} ,
\]
and $N_{\mathrm{traj}}$ is the number of quantum trajectories. 
The degree of mixedness of the reduced density matrix is quantified by the purity 
\[
\mathcal{P}_i={\rm Tr}(\hat{\rho}_i^2),
\]
whose deviation from unity measures the loss of coherence. As shown in Fig~\ref{Supp_Fig4}(b), $\mathcal{P}_i$ decreases slowly from unity at long time, but remains above the value $1/d$ (with $d=2S+1$) corresponding to the maximally mixed state. The deviation from this maximally mixed state increases with increasing spin magnitude $S$.
The von Neumann entropy of the reduced density matrix 
\begin{equation}
\mathcal{S}^{\mathrm{VN}}(t)=-{\rm Tr}\!\left[\hat{\rho}_i(t)\ln \hat{\rho}_i(t)\right]
\end{equation}
serves as an additional useful diagnostic to distinguish regular dynamics from chaotic behavior (as discussed in the main text). In the oscillatory regime, $\mathcal{S}^{\mathrm{VN}}(t)$ grows slowly with oscillations and does not exhibit sustained linear growth [Fig~\ref{Supp_Fig4}(c)], consistent with regular dynamical behavior.

Phase coherence is also reflected in fluctuations of the relative phase $\phi_i$ between the two wells. We find that $(\Delta\phi_i)^2$ increases slowly [Fig.~\ref{Supp_Fig4}(d)] and the growth rate decreases with increasing $S$, indicating that each condensate preserves a high degree of phase coherence over long times during the oscillatory motion. In contrast, the  chaotic regime exhibits reduced purity, rapid growth of entropy, and enhanced phase fluctuations [see Fig.~4(h,i) of the main text].

\section*{Supplementary note 4:\\ Steady-state chaos}
\label{Section3}
\begin{figure}[h]
\centering
\includegraphics[width=0.8\linewidth]{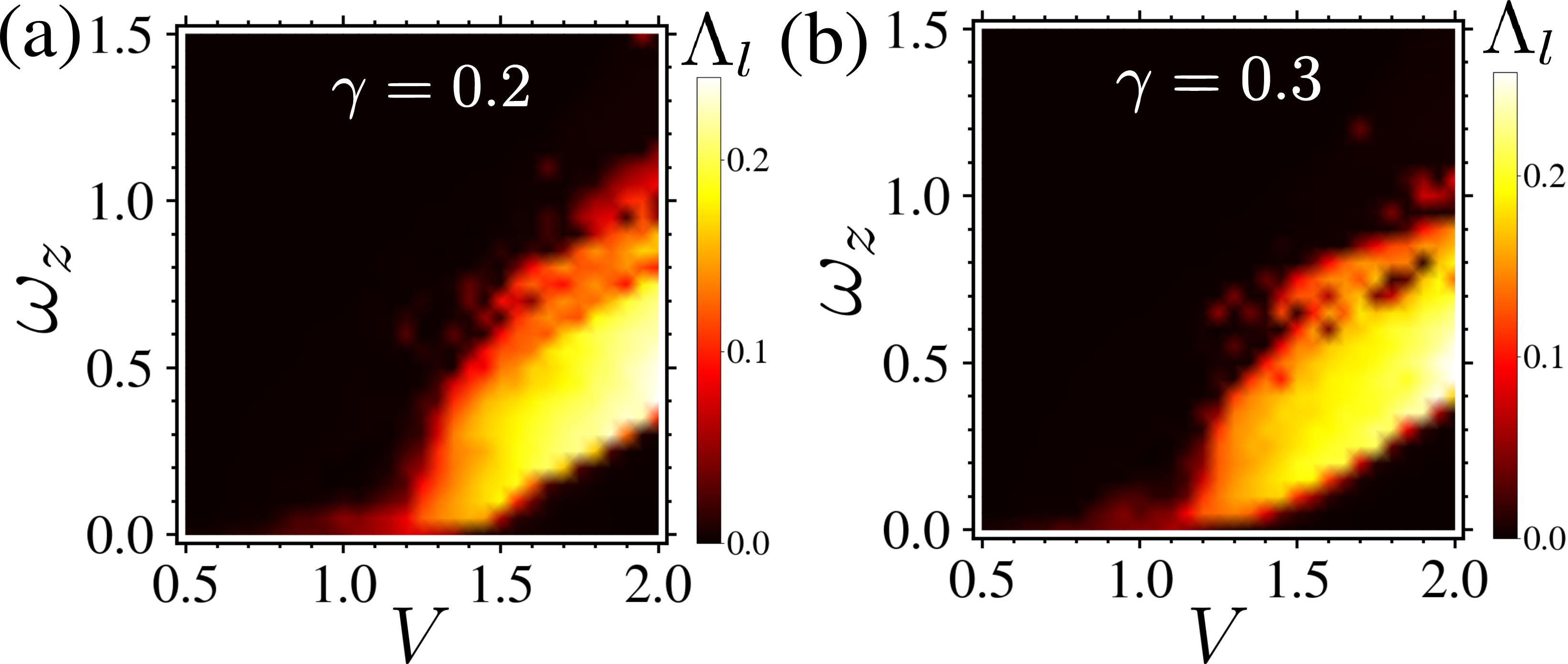}
\caption{{\bf Steady-state chaos.}
The regime of steady-state chaos identified by the positive value of the average Lyapunov exponent $\Lambda_l$ (color scale) in the $\omega_z$--$V$ plane for dissipation strengths (a) $\gamma=0.2$ and (b) $\gamma=0.3$.
}
\label{Supp_Fig3}
\end{figure}

In the presence of a tilt between the two wells of the BJJ, corresponding to the perturbation 
\[
\hat{\mathcal{H}}_{\rm P} = \frac{\omega_z}{2}\sum_i(\hat{n}_{i\rm L}-\hat{n}_{i\rm R}),
\]
steady-state chaos emerges within a finite region of $\omega_z$  (as discussed in the main text).
Classically, the degree of chaos is quantified by the Lyapunov exponent (LE) \cite{Strogatz}, which measures the sensitivity of initial conditions. To identify the region of steady-state chaos, we compute the mean Lyapunov exponent $\Lambda_l$, averaged over an ensemble of initial phase-space points.
Figure~\ref{Supp_Fig3} shows $\Lambda_\ell$ over the $\omega_z$--$V$ plane for fixed dissipation strength $\gamma$. The LE is positive within a finite region, revealing steady-state chaos even in the presence of dissipation. Moreover, the extent of the steady-state-chaos region changes only weakly as $\gamma$ is increased.

\section*{Supplementary note 5:\\ Spectral statistics of the Liouvillian} 

We now investigate quantum signatures of dissipative chaos via the spectral statistics of the Liouvillian, the generator of density-matrix evolution. The Liouvillian superoperator can be written as
\begin{eqnarray}
	\hat{\mathcal{L}} =\! -i[\hat{\mathcal{H}}\otimes\mathbb{I}-\mathbb{I}\otimes\hat{\mathcal{H}}^{\rm TR}]+ \sum_i \!\left(\hat{\mathcal{O}}_i\otimes \hat{\mathcal{O}}_i^*-\frac{1}{2}\hat{\mathcal{O}}_i^{\dagger}\hat{\mathcal{O}}_i\otimes \mathbb{I}-\frac{1}{2}\mathbb{I}\otimes \hat{\mathcal{O}}_i^{\rm TR}\hat{\mathcal{O}}_i^*\!\right),\quad
\end{eqnarray} 	
where the superscript $\mathrm{TR}$ denotes transposition and $\hat{\mathcal{O}}_i$ are the jump operators.

The isolated coupled-top model possesses a discrete symmetry under exchange of the two spins, $\vec{S}_1 \leftrightarrow \vec{S}_2$. This symmetry is represented by the operator $\hat{\Pi}$, which interchanges the indices of the basis states $\ket{m_{1z}, m_{2z}}$, where $m_{iz}$ are the eigenvalues of $\hat{S}_{iz}$, such that $[\hat{\mathcal{H}}, \hat{\Pi}] = 0$.
The corresponding superoperator  
\begin{eqnarray}
	\hat{\Pi}_s = \hat{\Pi} \otimes \hat{\Pi}^* ,
	\label{Exchange_Superoperator}
\end{eqnarray}
describes a weak exchange symmetry of Liouvillian, such that $[\hat{\mathcal{L}},\hat{\Pi}_s ] = 0$.
Consequently, the Liouvillian $\hat{\mathcal{L}}$ is block diagonal, with each block labeled by the eigenvalues $\Pi_s = \pm 1$. In the spectral analysis below, we consider eigenvalues within a fixed $\Pi_s$ sector.

{\bf Spacing statistics.}
To analyze spectral correlations of the Liouvillian spectrum, we study the distribution of the nearest-neighbor spacing $\delta$ (Euclidean distance between nearest eigenvalues in the complex plane).
From the raw data of the spectrum, we perform the unfolding procedure following~\cite{Prosen_PRL}. We also consider the complex spacing ratio
\begin{eqnarray}
	\xi_{\nu} = \frac{E_{\nu}^{\rm NN}-E_{\nu}}{E_{\nu}^{\rm NNN}-E_{\nu}} = r_{\nu} e^{i\theta_{\nu}},
\end{eqnarray}
where $E_{\nu}^{\rm NN}$ and $E_{\nu}^{\rm NNN}$ are the nearest and next-nearest neighbors of the complex eigenvalue $E_{\nu}$. 

\begin{figure}[h]
\centering
\includegraphics[width=0.9\linewidth]{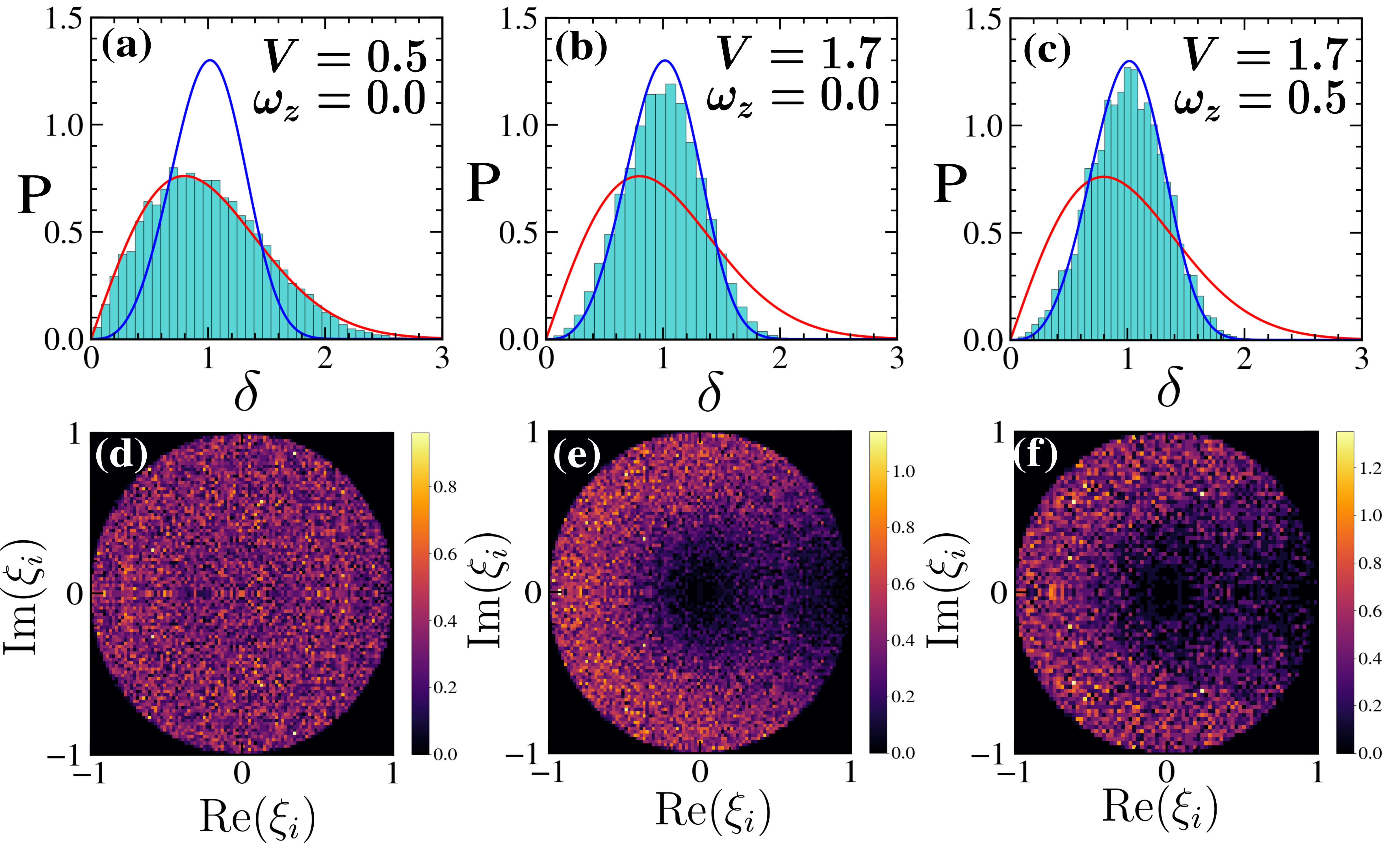}
\caption{
{\bf Spectral statistics of the Liouvillian superoperator of the open BJJ.} (a-c) Level spacing distributions P$(\delta)$ and (d-f) complex spacing ratio $\xi$ for the (a,d) oscillatory regime with $V=0.5,\omega_z=0.0$, (b,e) in the transient chaotic regime at $V=1.7,\omega_z=0.0$, and (c,f) in the regime of steady-state chaos with $V=1.7, \omega_z=0.5$. The red (blue) solid lines in  (a-c) represent the 2D-Poisson (Ginibre) distribution of the random matrix theory. Parameters: $S=8, \gamma=0.2$.
}
\label{Supp_Fig5}
\end{figure}

According to the Grobe-Haake-Sommers (GHS) conjecture \cite{Haake_GHS}, in the regular regime, the level spacing distribution follows the two-dimensional(2D) Poisson distribution with linear level repulsion, whereas, in the chaotic regime, it exhibits Ginibre correlation with cubic level repulsion \cite{Haake_GHS}.
In the oscillatory (regular) regime of the dissipative BJJ, the level spacing of Liouvillian level-spacing distribution [Fig.\ref{Supp_Fig5}(a)] agrees with the 2D-Poisson distribution, given by \cite{Prosen_PRX}
\[
{\rm P}_{\rm 2d-P}(\delta) = \frac{\pi}{2}\delta \exp(-\pi\delta^2/4),
\]
and the complex spacing ratio [Fig.\ref{Supp_Fig5}(d)] exhibits isotropic distribution with $\langle\cos\theta\rangle\approx0$.

In the steady-state chaotic regime, the level-spacing distribution reveals Ginibre correlations with ${\rm P}(\delta)\sim \delta^3$ for small $\delta$~\cite{Prosen_PRX} [Fig.~\ref{Supp_Fig5}(c)] and the complex spacing ratio exhibits a clear anisotropy quantified by $-\langle \cos\theta\rangle\approx 0.24$ [Fig.~\ref{Supp_Fig5}(f)], consistent with Ginibre statistics and in accordance with the GHS conjecture.

Surprisingly, Ginibre statistics is also observed in the transient-chaotic regime [Fig.~\ref{Supp_Fig5}(b,e)], even though the long-time dynamics becomes regular due to the convergence toward the stable attractor FP-III, thus violating the GHS conjecture. This illustrates that Liouvillian spectral statistics can diagnose short-time chaotic mixing but does not, by itself, distinguish transient chaos from steady-state chaos.

\bibliography{Supp_bibliography.bib}